\documentclass[twocolumn,pre,amsmath,amssymb,unsortedaddress,showpacs,nofootinbib]{revtex4-1}

\usepackage{latexsym,epsfig,graphicx}% Include figure files
\usepackage{dcolumn}% Align table columns on decimal point
\usepackage{bm}% bold math
\usepackage{color} %coloring for comments
\usepackage{graphicx}
\usepackage{subfigure}
\usepackage{accents}
\usepackage{tikz}

\usepackage[colorlinks,urlcolor=blue,citecolor=blue]{hyperref}

\usetikzlibrary{arrows,snakes,automata,shadows,backgrounds,positioning}
\newcommand{\ubar}[1]{\underaccent{\bar}{#1}}

\allowdisplaybreaks

\begin{document}

\title{Derivation of matrix product states for the Heisenberg spin chain \\with open boundary conditions}

\author{Zhongtao Mei}
\author{C. J. Bolech}
\affiliation {Department of Physics, University of Cincinnati, Cincinnati, Ohio 45221-0011, USA}
\date{\today}

\begin{abstract}
Using the algebraic Bethe ansatz, we derive a matrix product representation of the exact Bethe-ansatz states of the six-vertex Heisenberg chain (either XXX or XXZ and spin-$\frac{1}{2}$) with open boundary conditions. In this representation, the components of the Bethe eigenstates are expressed as traces of products of matrices that act on a tensor product of auxiliary spaces. As compared to the matrix product states of the same Heisenberg chain but with periodic boundary conditions, the dimension of the exact auxiliary matrices is enlarged as if the conserved number of spin-flips considered would have been doubled. This result is generic for any non-nested integrable model, as is clear from our derivation, and we further show this by providing an additional example of the same matrix product state construction for a well known model of a gas of interacting bosons. Counter\-intuitively, the matrices do not depend on the spatial coordinate despite the open boundaries, and thus they suggest generic ways of exploiting (emergent) translational invariance both for finite size and in the thermodynamic limit.
\end{abstract}
\maketitle

\section{Background and Motivation}\label{sec:introduction}

Spin models are the simplest representation for the description of magnetic materials, and for a long time they have played a central role in the study of the statistical and thermodynamical properties of that important class of systems. Chains of spins are used to study the one-dimensional case, which is relatively simple for classical spins but it already shows a rich algebraic structure when quantum spins, obeying a Heisenberg algebra, are introduced. The simplest of quantum spin chains, the XXX spin-$\frac{1}{2}$ Heisenberg model, is the one that was famously solved by Bethe in his 1931 landmark paper in which he introduces the key ideas that were eventually generalized into what is now known as the Bethe Ansatz~\cite{Bethe1931} ---a method of choice for the study of low-dimensional integrable systems. The introduction of axial anisotropy, breaking the full rotational symmetry in spin space, yields the so-called XXZ Heisenberg model. Remarkably, it can be seen via a standard quantum-classical correspondence that these spin chains are equivalent, for the case of spin-$\frac{1}{2}$, to a so-called six-vertex model ---the two-dimensional generalization of a model used to study the phases of water ice~\cite{Bernal1933,Pauling1935} (notice it was put forward at about the same time as when Bethe was working on magnetism). Each oxygen in the ice crystal is surrounded by four other oxygen atoms in a tetrahedral configuration, and the four intervening hydrogen bonds are such that while two protons are strongly connected to that oxygen, the other two are weakly so. That way each individual-oxygen environment resembles the configuration of a water molecule. A planar model of this on a square lattice, sometimes referred to as square ice, was famously studied using Bethe's ansatz by Lieb~\cite{Lieb1967a,Lieb1967b}, generalized by Sutherland~\cite{Sutherland1967}, and further generalized by Baxter (who also extended the solution to eight-vertex configurations, which are equivalent to the fully anisotropic XYZ chain), and it serves as a cornerstone for the theory of integrable systems \cite{Baxter}.

Despite the integrability of these systems, some aspects of their physics, for example the spin-spin correlations and dynamical properties, can be notoriously difficult to compute. Moreover, the physically motivated introduction of variants in the models will most often spoil their integrability. There is thus a clear need for computational methods that can enhance our calculation abilities for these systems and build on the integrable cases as test benches. In the case of spin chains and planar ice-type models, a momentous advance took place in the 1990s with the development of the density matrix renormalization group (DMRG) \cite{White1992,White1993} and the related (corner) transfer-matrix renormalization group ((C)TMRG) \cite{Nishino1996}. The great accuracy and versatility of these algorithms triggered a large amount of activity, not only in terms of applications but also to understand better the foundations of their remarkable performance. Some of the latter activity and insights came from the analysis of DMRG and (C)TMRG from the modern point of view of entanglement and quantum information theory. It was first pointed out that these methods can be reinterpreted variation\-ally as optimizations in a space of matrix products \cite{Ostlund1995}. Thus, in recent years, such matrix product states (MPSs) have attracted a renewed interest in both fields of quantum statistical physics and quantum information science~\cite{VerstraeteMurgCirac2008,garcia2004,verstraete2004,popp2005}. It was already known by then that for the case of the Affleck-Kennedy-Lieb-Tasaki (AKLT) model~\cite{Affleck1987}, the exact ground state can be written as a MPS, but the more intricate connection between MPS and Bethe-ansatz integrable models had not yet been developed. Independently of that, the connection between MPS methods and DMRG has since been firmly established and led to practical advances~\cite{Schollwock2011, SchollwockRevModPhys2005}. Furthermore, MPSs are seen as essential tools for the modern and expanding field of tensor networks~\cite{Orus2014}.

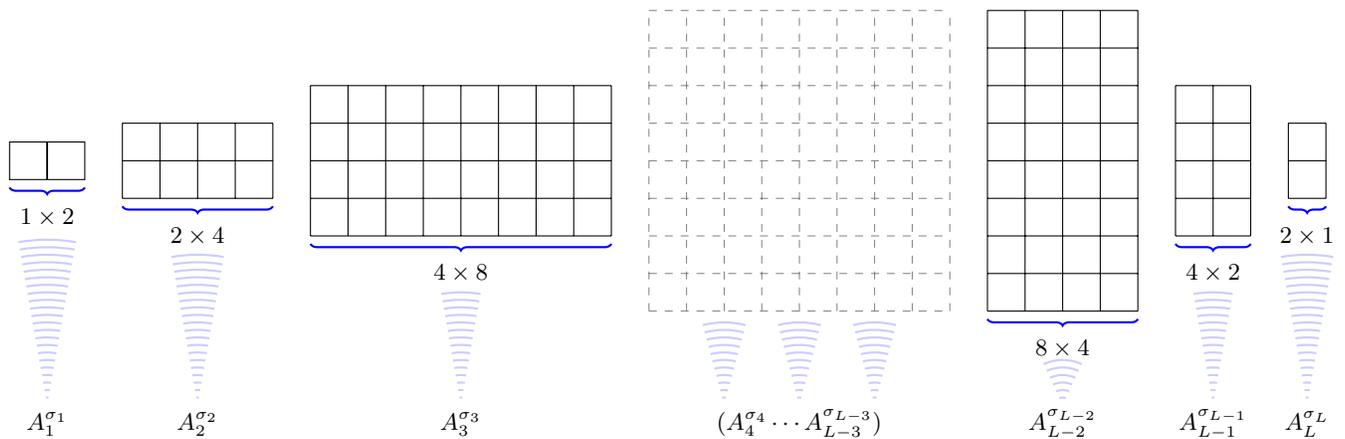
\begin{figure*}
\begin{tikzpicture}[scale=0.5, set style={{help lines}+=[dashed]}]
\draw (0,3.5) rectangle +(1,1) (1,3.5) rectangle +(1,1);
\draw (3,3) grid +(4,2);
\draw (8,2) grid +(8,4);
\draw[style=help lines] (17,0) grid +(8,8);
\draw (26,0) grid +(4,8);
\draw (31,2) grid +(2,4);
\draw (34,3) grid +(1,2);
\draw[snake=brace, raise snake=3pt, thick, blue] (2,3.5) -- +(-2,0);
\draw[snake=brace, raise snake=3pt, thick, blue] (7,3) -- +(-4,0);
\draw[snake=brace, raise snake=3pt, thick, blue] (16,2) -- +(-8,0);
\draw[snake=brace, raise snake=3pt, thick, blue] (30,0) -- +(-4,0);
\draw[snake=brace, raise snake=3pt, thick, blue] (33,2) -- +(-2,0);
\draw[snake=brace, raise snake=3pt, thick, blue] (35,3) -- +(-1,0);
\node at (1,2.5) {$1\times2$};
\node at (5,2) {$2\times4$};
\node at (12,1) {$4\times8$};
\node at (28,-1) {$8\times4$};
\node at (32,1) {$4\times2$};
\node at (34.5,2) {$2\times1$};
\node at (1,-3) {$A_{1}^{\sigma_{1}}$};
\node at (5,-3) {$A_{2}^{\sigma_{2}}$};
\node at (12,-3) {$A_{3}^{\sigma_{3}}$};
\node at (21,-3) {$(A_{4}^{\sigma_{4}} \cdots A_{L-3}^{\sigma_{L-3}})$};
\node at (28,-3) {$A_{L-2}^{\sigma_{L-2}}$};
\node at (32,-3) {$A_{L-1}^{\sigma_{L-1}}$};
\node at (34.5,-3) {$A_{L}^{\sigma_{L}}$};
\draw [snake=expanding waves, segment length=1mm, segment angle=10, thick, blue!20] (1,-2.5) -- (1,2);
\draw [snake=expanding waves, segment length=1mm, segment angle=10, thick, blue!20] (5,-2.5) -- (5,1.5);
\draw [snake=expanding waves, segment length=1mm, segment angle=10, thick, blue!20] (12,-2.5) -- (12,0.5);
\draw [snake=expanding waves, segment length=1mm, segment angle=15, thick, blue!20] (19,-2.5) -- (19,-0.3);
\draw [snake=expanding waves, segment length=1mm, segment angle=15, thick, blue!20] (21,-2.5) -- (21,-0.3);
\draw [snake=expanding waves, segment length=1mm, segment angle=15, thick, blue!20] (23,-2.5) -- (23,-0.3);
\draw [snake=expanding waves, segment length=1mm, segment angle=25, thick, blue!20] (28,-2.5) -- (28,-1.3);
\draw [snake=expanding waves, segment length=1mm, segment angle=10, thick, blue!20] (32,-2.5) -- (32,0.5);
\draw [snake=expanding waves, segment length=1mm, segment angle=10, thick, blue!20] (34.5,-2.5) -- (34.5,1.5);
\end{tikzpicture}
\caption{\label{fig:MPSwOBC} Example of a typical MPS coefficient generated by an iterative singular-value decomposition. In the generic case of all non-zero singular values, the $A_i^{\sigma_i}$ matrix sizes telescope as $(1 \times d), (d \times d^2), \dots, (d^{[L/2]} \times d^{[L/2]}), \dots, (d^2 \times d), (d \times 1)$; where at the center of the chain there will be one single square matrix if the number of sites, $L$, is odd or a pair of matrices of transposed dimensions if $L$ is even \cite{Schollwock2011} (the square brackets denote the integer-part of $L/2$). Here $d$ is the dimension of the local Hilbert space for each lattice site, ---we have $d=2$ for the case of the spin-$\frac{1}{2}$ chains that we study and as depicted in the figure. The dashed matrix in the center represents the product of all the central matrices that, in principle, telescope to very large sizes. In practical implementations these need to be truncated and are replaced by uniformly square matrices of size $D \times D$, where D is called the \textit{bond dimension}. This makes the practical OBC-MPS ansatz in the bulk of the chain look similar to that one used for PBC discussed in the text.} 
\end{figure*}

More recently, a matrix product ansatz proposed by Alcaraz and Lazo~\cite{Alcaraz2004,Alcaraz2006} gave the exact eigenstates of integrable spin chains expressed as MPSs. At roughly the same time, a MPS for the so-called asymmetric exclusion process (ASEP) was derived from the algebraic Bethe ansatz\footnote{Notice that direct studies using matrix formulations but no Bethe Ansatz predate this, e.g.~\cite{Derrida1993}, and continue largely in parallel with occasional connections; for recent examples, see Refs.~\onlinecite{Aneva2016,Crampe2016}.} (ABA)~\cite{GolinelliMallick2006} ---which is an elegant method for solving the eigenvalue problem of quantum integrable models developed in the late 1970s~\cite{Nepomechie1999,Korepin1993}. Along the same lines as for the ASEP case, a MPS form for the eigenstates of the six-vertex spin chain (referring indistinguishably to the XXX or XXZ cases and spin-$\frac{1}{2}$) was derived using the ABA~\cite{Katsura2010}, and the authors also showed that the MPSs they obtained were equivalent to those proposed by Alcaraz and Lazo. Using similar ideas, the same authors derived a continuous matrix product state (cMPS) for the Lieb-Liniger model (i.e.~a gas of bosons interacting via a short-range contact potential~\cite{LiebLiniger1963}) from the ABA~\cite{Maruyama2010}. For all these works about the ASEP, the integrable spin chains, and the Lieb-Liniger model, only periodic boundary conditions (PBC) were considered. Until now, there was no similar work extending those results to the ubiquitous case of open boundary conditions (OBC).

%Motivate different BCs
The importance of boundary conditions, however, should not be overlooked. First, the meanings of the boundary conditions for the spin chain and for the related vertex model need to be differentiated. This is especially true since in the case of the six-vertex ice model, it has been pointed out that, counter\-intuitively, the bulk free energy can depend on the boundary conditions \cite{Korepin2000,ZinnJustin2000,Cugliandolo2015} ---this, of course, requires the use of special fixed boundary conditions, such as the so called domain-wall boundary conditions (DWBC); the non-fixed common choices of open and periodic conditions are indeed equivalent in the thermodynamic limit~\cite{Brascamp1973}. Within the ABA construction, the six-vertex Heisenberg spin chain with PBC is mapped into a six-vertex ice-type model in the space of monodromies with fixed boundary conditions (in particular, DWBC are relevant~\cite{Katsura2010}). So the boundary conditions of the spin chain and of the related vertex model are different. If one is instead interested in having OBC for the spin chain, the space of monodromies is different and has a rather more complicated structure ---we shall be discussing it in detail below during our calculation. This takes us to a second point: what is the motivation for considering OBC? The answer comes from an interest in the properties of the DMRG algorithm as applied to spin chains. It was observed early on that the accuracy and speed of convergence of DMRG were much better in the case of OBC as compared with PBC. It was subsequently understood that the root cause of this is the structure of the MPSs generated by the algorithm. These have the form $\sum_{\{\sigma\}}\,A^{\sigma_1}_1\cdots A^{\sigma_L}_L\,\left|\sigma_1\dots\sigma_L\right>$, where the matrices are in general rectangular and telescope both ways in size so that their product is ultimately a scalar quantity (for more details see Fig.~\ref{fig:MPSwOBC}). Without truncation of the matrix sizes, this can in principle represent any arbitrary quantum state. With truncation, it works well for OBC, but it has the problem that the first-few and last-few matrices are too restricted to fully capture the entanglement between the first-few and last-few spins ---which with PBC is large since they actually become close neighbors by virtue of closing the chain on itself. The exact construction overcomes this problem by allowing exponentially large (in the number of sites) matrices for the central portion of the chain, but that is computationally impracticable. Thus, in the PBC case, it is more natural to postulate MPSs of the form $\sum_{\{\sigma\}}\,\mathrm{Tr}\left[A^{\sigma_1}_1\cdots A^{\sigma_L}_L\right]\,\left|\sigma_1\dots\sigma_L\right>$, where all the matrices can now be taken to be square and equal in size. The effect of truncating the matrices is now distributed uniformly along the chain and is less detrimental. It was indeed shown that, in a variational MPS-based reformulation of DMRG, the algorithmic performance with either type of boundary conditions becomes comparable if the matching type of MPS is used~\cite{VerstraetePorrasCirac2004}. Interestingly, the ABA construction of MPSs for the periodic six-vertex spin chain naturally gives a traced-product of matrices in auxiliary space. The study of what happens for the open six-vertex spin chain will be the main focus of the present work.

In this paper, we shall construct MPSs for the six-vertex Heisenberg spin chain with OBC by using the ABA. For the MPSs of ditto system but with PBC~\cite{Katsura2010}, the dimension of the matrices is $2^{n}$, where $n$ is the number of spin-flips present in the state. For OBC, we shall see that one cannot construct a MPS with the dimension of the matrices being $2^{n}$, but one can obtain a MPS with matrices of dimension $2^{2n}$ instead. The matrices of the MPS will not be given in closed form, but they can be obtained for any case of interest by using a set of recursion relations that will be derived. We shall also extend the same ideas to construct a MPS for a lattice-regularized version of the Lieb-Liniger model with OBC. To obtain a closed-form cMPS similar to the one in~\cite{VerstraeteCirac2010} for the Lieb-Liniger model with OBC, one has to solve explicitly the recursion relations of the matrices, which shall remain as an open problem. 

The paper is outlined as follows. In section~\ref{sec:ABA of XXZ}, we present the ABA method for solving the eigenvalue problem of the six-vertex spin chain with OBC. In section~\ref{sec:MPS for XXZ}, we derive the MPS representations for the Bethe eigenstates from the ABA. In section~\ref{sec:conclusion}, we conlcude and summarize our work. The MPS representations for a lattice-regularized version of the Lieb-Liniger model with OBC are given in appendix~\ref{sec:ABA of Lieb-Liniger}.

\section{Algebraic Bethe ansatz for the six-vertex spin chain with OBC}\label{sec:ABA of XXZ}
The six-vertex Heisenberg spin chain with OBC is described by the following Hamiltonian:
\begin{equation}\label{eq:HamXXZ}
H = \sum_{i=1}^{L-1}\left\{ \sigma_{i}^{x}\sigma_{i+1}^{x}+\sigma_{i}^{y}\sigma_{i+1}^{y}
+\Delta(\sigma_{i}^{z}\sigma_{i+1}^{z}-1)\right\}-\Delta\,,
\end{equation}
where $L$ denotes the total number of sites, $\sigma_{i}^{\alpha}$ (with $\alpha=x, y, z$) are the Pauli matrices for the spin on the $i$th site, and $\Delta$ is the anisotropy parameter (which is $1$ for the XXX case). The eigenstates of this model can be constructed using the ABA~\cite{Korepin1993,Nepomechie1999}; we shall now briefly outline this construction. The central object of the ABA is the quantum $R$-matrix which is the solution of the Yang-Baxter equation,
\begin{equation*}
R_{ab}(\lambda-\lambda') R_{ac}(\lambda) R_{bc}(\lambda') = 
R_{bc}(\lambda') R_{ac}(\lambda) R_{ab}(\lambda-\lambda')\,.
\end{equation*}
 For the six-vertex spin chain, the $R$-matrix acting on the space $V_{a} \otimes V_{b}$ has only six non-zero entries and is given by
\begin{equation}
R_{ab}(\lambda) = \begin{pmatrix}1 & 0 & 0 & 0\\
0 & \frac{\sinh\lambda}{\sinh(\lambda+\eta)} & \frac{\sinh\eta}{\sinh(\lambda+\eta)} & 0\\
0 & \frac{\sinh\eta}{\sinh(\lambda+\eta)} & \frac{\sinh\lambda}{\sinh(\lambda+\eta)} & 0\\
0 & 0 & 0 & 1
\end{pmatrix}_{ab}\,, 
\end{equation} 
where $\eta$ is defined by $\Delta=\cosh\eta$ (notice that the limit $\eta\to0$ for the $R$-matrix is not well posed within the parametrization we are using for the ABA construction, and that case is better handled separately,\footnote{The $R$-matrix for the XXX case is \textit{rational} instead of \textit{trigonometric} and is given by
%\begin{equation*}
%R_{ab}(\lambda) = \begin{pmatrix}1 & 0 & 0 & 0\\
%0 & \frac{\lambda}{\lambda+\eta} & \frac{\eta}{\lambda+\eta} & 0\\
%0 & \frac{\eta}{\lambda+\eta} & \frac{\lambda}{\lambda+\eta} & 0\\
%0 & 0 & 0 & 1
%\end{pmatrix}_{ab}\,.  
%\end{equation*}
the same expression with the replacement $\sinh(\lambda)\to\lambda$ everywhere and the limit substitution $\eta\to{0+i}$. Alternatively, one can rescale $\lambda\to\lambda\epsilon$ and substitute $\eta\to i\epsilon$ and then take the $\epsilon\to0$ limit (so that $\Delta\to1$).} 
although all the formal constructs are parallel to the XXZ case). 
Next, we introduce the quantum $L$-operator represented by a matrix acting on the tensor product of two two-dimensional vector spaces $V_{0} \otimes V_{i}$. The two-dimensional auxiliary space is denoted by $V_{0}$ (or also $V_{a}$ or $V_{b}$ when we need two of them below), and the physical Hilbert space at the $i$th site is denoted by $V_{i}$. For the Heisenberg spin-$\frac{1}{2}$ chain we have
\begin{equation} 
L_{0i} (\lambda) = R_{0i}(\lambda - \frac{\eta}{2})\,.
\end{equation}
The following intertwining relation can be shown as a direct consequence of the Yang-Baxter equation
\begin{equation}\label{eq:YBE}
R_{ab}(\lambda-\mu) L_{ai}(\lambda) L_{bi}(\mu) = L_{bi}(\mu) L_{ai}(\lambda) R_{ab}(\lambda-\mu)\,.
\end{equation}
Here, $R_{ab}$ acts on the space $V_{a} \otimes V_{b}$, and $L_{ai}$ acts on $V_{a} \otimes V_{i}$. $V_{a}$ and $V_{b}$ are auxiliary spaces, and $V_{i}$ is the physical space at site $i$.

The construction so far did not involve the boundary conditions, but let us now focus on the OBC case. Following Sklyanin, we can consider a looped monodromy matrix~\cite{Sklyanin1988}
\begin{equation}\label{eq:Tmat} 
T_{0}(\lambda)=L_{01}(\lambda)\cdots L_{0L}(\lambda)L_{0L}(\lambda)\cdots L_{01}(\lambda)\,.
\end{equation}
In the space $V_{0}$, the monodromy matrix can be represented as a $2 \times 2$ matrix
\begin{equation}
T_{0}(\lambda) = \begin{pmatrix} A(\lambda) & B(\lambda) \\
C(\lambda) & D(\lambda)
\end{pmatrix}_{0}\,,  
\end{equation} 
where the matrix elements $A(\lambda)$, $B(\lambda)$, $C(\lambda)$ and $D(\lambda)$ are themselves operators acting on the total Hilbert space of the chain $V_{1} \otimes V_{2} \otimes \cdots \otimes V_{L}$. Using Eq.~(\ref{eq:YBE}), we can obtain the following relation for the monodromy matrix
\begin{multline}\label{eq:YBE2}
R_{ab}(\lambda-\mu) T_{a}(\lambda) R_{ab}(\lambda+\mu-\eta) T_{b}(\mu) = \\ T_{b}(\mu) R_{ab}(\lambda+\mu-\eta) T_{a}(\lambda) R_{ab}(\lambda-\mu)\,.
\end{multline}
The commutation relations among $A$, $B$, $C$ and $D$ can be obtained from this relation. 

Taking the trace of the monodromy matrix over $V_{0}$, we obtain a one-parameter family of transfer matrices acting on the total physical Hilbert space $V_{1} \otimes V_{2} \otimes \cdots \otimes V_{L}$:
\begin{equation}
t(\lambda)= \textrm{Tr}_{V_{0}}T_{0}(\lambda)=A(\lambda)+D(\lambda)\,.
\end{equation} 
One can show that $[t(\lambda), t(\mu)] = 0$~\cite{Sklyanin1988}. Furthermore, the Hamiltonian can be recovered from the logarithmic derivative of $t(\lambda)$
\begin{equation}
H=\sinh\eta \left[ \frac{\partial}{\partial\lambda}\log t(\lambda) \right]_{\lambda=\eta/2}\,.
\end{equation} 

Since the Hamiltonian commutes with the monodromy matrix, we can construct simultaneous eigenstates of both $H$ and $t(\lambda)$. Such an eigenstate of $t(\lambda)$ is given by
\begin{equation}
\left|\lambda_{1},\lambda_{2},\cdots,\lambda_{n}\right\rangle =B(\lambda_{1})B(\lambda_{2})\cdots B(\lambda_{n})\left|\omega\right\rangle\,, 
\end{equation}
where $\left|\omega\right\rangle$ denotes the standard reference ferromagnetic state, i.e.~$\left|\omega\right\rangle = \left|\uparrow\right\rangle _{1} \left|\uparrow\right\rangle _{2} \cdots \left|\uparrow\right\rangle _{L}$, and $n$ denotes the number of down spins. We call the above state a Bethe state. Since one can show from Eq.~(\ref{eq:YBE2}) that the $B(\lambda_{i})$'s commute with each other, this state is invariant under permutations of $\lambda_{i}$'s. The latter have to be non-repeating and obey the OBC Bethe ansatz equations, which are given by
\begin{multline}\label{eq:BAE_XXZ}
\prod_{\substack{j=1\\ j\neq i}}^{n}
\frac{\sinh(\lambda_{i}+\lambda_{j}+\eta)\sinh(\lambda_{i}-\lambda_{j}+\eta)}{\sinh(\lambda_{i}+\lambda_{j}-\eta)\sinh(\lambda_{i}-\lambda_{j}-\eta)} = \\
\frac{\cosh^{2}(\lambda_{i}-\frac{\eta}{2})\sinh^{2L}(\lambda_{i}+\frac{\eta}{2})}{\cosh^{2}(\lambda_{i}+\frac{\eta}{2})\sinh^{2L}(\lambda_{i}-\frac{\eta}{2})}\,, \\
i=1,2,\dots n\,.
\end{multline} 

The eigen\-energy corresponding to $ \left|\lambda_{1},\lambda_{2},\cdots,\lambda_{n}\right\rangle $ is given by
\begin{equation}\label{eq:E_XXZ}
E(\lambda_{1},\lambda_{2},\cdots,\lambda_{n}) = \sum_{i=1}^{n} 
\frac{2 \sinh^{2}(\eta)}{\sinh(\lambda_{i} + \frac{\eta}{2}) \sinh(\lambda_{i} - \frac{\eta}{2})} - \Delta\,.
\end{equation} 

\section{Exact matrix product states for the six-vertex spin chain with OBC}\label{sec:MPS for XXZ}

In the previous section, we outlined the construction of the eigenstates of $H$ using the ABA. In this section, we derive MPS representations for those eigenstates. 

Using
\begin{eqnarray}
B(\lambda)= \hspace{0.001 in} _{0}\!\left\langle \uparrow\right|T_{0}(\lambda)\left|\downarrow\right\rangle _{0},
\end{eqnarray}
we can write
\begin{multline}\label{eq:MPS1}
\left|\lambda_{1},\lambda_{2},\cdots,\lambda_{n}\right\rangle = \hspace{0.001 in} _{\bar{n}}\!\left\langle \uparrow\right| \cdots \hspace{0.001 in} _{\bar{2}}\!\left\langle \uparrow\right| \hspace{0.001 in} _{\bar{1}}\!\left\langle \uparrow\right| \cdot \\ T_{\bar{n}}(\lambda_{n})\cdots T_{\bar{2}}(\lambda_{2})T_{\bar{1}}(\lambda_{1}) \cdot \left|\downarrow\right\rangle _{\bar{n}}\cdots\left|\downarrow\right\rangle _{\bar{2}}\left|\downarrow\right\rangle _{\bar{1}}\cdot\left|\omega\right\rangle\,. 
 \end{multline}
To derive a MPS for this Bethe state, we need to reorganize the $L$-operators in that product of monodromy matrices. All of those belonging to the same physical Hilbert space should be brought together. Let us illustrate this by a simple example with $L=2$, $n=2$ and PBC. We have $T_{\bar{2}}T_{\bar{1}}=(L_{\bar{2}1}L_{\bar{2}2})\cdot(L_{\bar{1}1}L_{\bar{1}2})$, of which the spectral parameters are omitted for brevity. Then we can easily reorder them to obtain $T_{\bar{2}}T_{\bar{1}}=(L_{\bar{2}1}L_{\bar{1}1})\cdot(L_{\bar{2}2}L_{\bar{1}2})$. This was a crucial step to derive a MPS for the Heisenberg spin chain with PBC~\cite{Katsura2010}. For the same example but with OBC, we have $T_{\bar{2}}T_{\bar{1}}=(L_{\bar{2}1}L_{\bar{2}2}L_{\bar{2}2}L_{\bar{2}1})\cdot(L_{\bar{1}1}L_{\bar{1}2}L_{\bar{1}2}L_{\bar{1}1})$ and it is clear that we cannot put the $L$-operators with the same physical-space subscript (say $1$ or $2$ in this example) together by means of a simple reordering. 

To overcome this difficulty, we rewrite Eq.~(\ref{eq:MPS1}) as
\begin{multline}
\left|\lambda_{1},\lambda_{2},\cdots,\lambda_{n}\right\rangle =[\textrm{Tr}_{V_{\bar{n}}}(Q_{\bar{n}}T_{\bar{n}}(\lambda_{n}))]\cdots \\  [\textrm{Tr}_{V_{\bar{2}}}(Q_{\bar{2}}T_{\bar{2}}(\lambda_{2}))][\textrm{Tr}_{V_{\bar{1}}}(Q_{\bar{1}}T_{\bar{1}}(\lambda_{1}))]\left|\omega\right\rangle\,, 
\end{multline}
where we introduced the definition
\begin{equation}\label{eq:Qmat}
Q_{\bar{i}} = \left|\downarrow\right\rangle _{\bar{i}}\cdot \hspace{0.001 in} _{\bar{i}}\!\left\langle \uparrow\right| = \begin{pmatrix} 0 & 0 \\
1 & 0
\end{pmatrix}_{\bar{i}}\,. 
\end{equation}
Now we focus on the structure of an individual OBC looped monodromy matrix, $T_{\bar{i}}(\lambda_{i})$, as given in Eq.~(\ref{eq:Tmat}).
%\begin{eqnarray}
%T_{\bar{i}}(\lambda_{i})=L_{\bar{i}1}(\lambda_{i})\cdots L_{\bar{i}L}(\lambda_{i})L_{\bar{i}L}(\lambda_{i})\cdots L_{\bar{i}1}(\lambda_{i}).
%\end{eqnarray}
What we need to do is to change $(L_{\bar{i}1}\cdots L_{\bar{i}L})\cdot (L_{\bar{i}L}  \cdots L_{\bar{i}1})$ to something similar to $(L_{\bar{i}1}\cdots L_{\bar{i}L})\cdot (L_{\bar{i}1}\cdots L_{\bar{i}L})$. To achieve this, we denote
\begin{equation}
L_{\bar{i}1}(\lambda_{i})\cdots L_{\bar{i}L}(\lambda_{i})=\begin{pmatrix}E & F\\
G & H
\end{pmatrix}_{\bar{i}} \equiv M_{\bar{i}}\,,
\end{equation}
where $E$, $F$, $G$ and $H$ are matrices acting on the total Hilbert space $V_{1} \otimes V_{2} \otimes \cdots \otimes V_{L}$. We also similarly denote
\begin{equation}
L_{\bar{i}L}(\lambda_{i})\cdots L_{\bar{i}1}(\lambda_{i})=\begin{pmatrix}U & V\\
X & Y
\end{pmatrix}_{\bar{i}} \equiv N_{\bar{i}}\,.
\end{equation} 
\begin{figure}[t]
%{\Large (a)}\\
\begin{tikzpicture}
\draw[very thick,rounded corners=0.2cm,shorten >=2pt]
(-4,1)--(0,1)--(0,-0.5);
\draw[<-,very thick,rounded corners=0.2cm,shorten >=2pt]
(0,0)--(0,-1)--(4,-1);
\draw[->,rounded corners=0.2cm,shorten >=2pt]
(3.5,-2)--(3.5,-0.6)--(2.6,-0.6)  (2.4,-0.6)--(0.6,-0.6)  
(0.4,-0.6)--(0.1,-0.6)  (-0.1,-0.6)--(-3.5,-0.6)--(-3.5,2);
\draw[->,rounded corners=0.2cm,shorten >=2pt]
(2.5,-2)--(2.5,-0.2)--(0.6,-0.2)  (0.4,-0.2)--(0.1,-0.2)  (-0.1,-0.2)--(-2.5,-0.2)--(-2.5,2);
\draw[->,rounded corners=0.2cm,shorten >=2pt]
(0.5,-2)--(0.5,0.6)--(0.1,0.6)  (-0.1,0.6)--(-0.5,0.6)--(-0.5,2);
\draw[right] (4,-1) node(bar_i){$\bar{i}$};
\fill (4,-1) circle (0.075cm) (-4,1) circle (0.075cm);
\draw[below] (3.5,-2) node(1){$1$};
\draw[below] (2.5,-2) node(2){$2$};
\draw[below] (1.5,-2) node(3){$\cdots$};
\draw[below] (0.5,-2) node(L){$L$};
\end{tikzpicture}
\vskip .5cm
%{\Large (b)}\\
\begin{tikzpicture}
\draw[thick,rounded corners=0.2cm, color=green!80!black] (-0.3,1.2) rectangle (0.8,-1.2);
\draw[->,very thick,rounded corners=0.2cm,shorten >=2pt]
(-4,1)--(0,1)--(0,-0.5);
\draw[->,very thick,rounded corners=0.2cm,shorten >=2pt]
(1,1)--(0.5,1)--(0.5,0.3);
\draw[<-,very thick,rounded corners=0.2cm,shorten >=2pt]
(0,0.5)--(0,-1)--(-4,-1);
\draw[<-,very thick,rounded corners=0.2cm,shorten >=2pt]
(0.5,-0.6)--(0.5,0.6)
(0.5,-0.3)--(0.5,-1)--(1,-1);
\draw[->,rounded corners=0.2cm,shorten >=2pt]
(-3.5,-2)--(-3.5,2);
\draw[->,rounded corners=0.2cm,shorten >=2pt]
(-2.5,-2)--(-2.5,2);
\draw[->,rounded corners=0.2cm,shorten >=2pt]
(-0.5,-2)--(-0.5,2);
\draw[right] (0,0) node(Q){\large $\tilde{Q}$};
\draw[right] (1,1) node(bar_i){$\bar{i}$};
\draw[right] (1,-1) node(ubar_i){$\ubar{i}$};
\fill (-4,-1) circle (0.075cm) (-4,1) circle (0.075cm);
\fill (1,-1) circle (0.075cm) (1,1) circle (0.075cm);
\draw[below] (-3.5,-2) node(1){$1$};
\draw[below] (-2.5,-2) node(2){$2$};
\draw[below] (-1.5,-2) node(3){$\cdots$};
\draw[below] (-0.5,-2) node(L){$L$};
\end{tikzpicture}
\caption{\label{fig:Tcartoon} Schematic depiction of the traced looped monodromy matrix used to define the transfer matrix in the formulation of ABA with OBC. The vertical lines indicate the local physical Hilbert spaces for the different spins along the chain and are labeled with the site index. The thicker horizontal line correspond to the auxiliary space used in the definition of the monodromy matrix, the ending dots are the points contracted when tracing over the auxiliary space to obtain the transfer matrix. Each crossing of a physical and an auxiliary space corresponds to an $L$-operator insertion, and the adjoining lines correspond to the respective index contractions. The first drawing (top) depicts the standard formulation using a single auxiliary space. The second drawing (bottom) shows how the vertical lines can be made into straight lines by spitting the auxiliary space into two also straight lines and reversing the order of the products in the lower one via matrix transpositions. The two-auxiliary-space formulation connects to the one-auxiliary-space one via the identification of a tensor $\tilde{Q}$ (green box) that merges the two spaces, see Eq.~(\ref{eq:tildeQ}), in a way that naturally connects with the tracing procedure. The matrix $Q$ was not included in the graphical representation; it can be added when closing the trace and combined with $\tilde{Q}$ into $\mathcal{Q}$; cf.~Figs.~\ref{fig:OBCnetwork} and \ref{fig:PBCnetwork}.}
\end{figure}
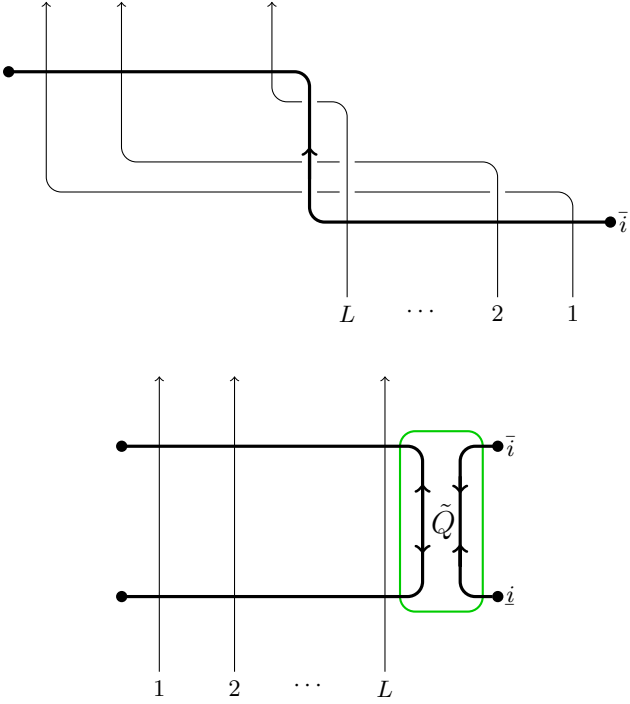
Next we introduce an additional auxiliary space $V_{\ubar{i}}$ and we want to find a $4 \times 4$ matrix $\mathcal{Q}_{\bar{i}\ubar{i}}$ such that
\begin{eqnarray}
\textrm{Tr}_{V_{\bar{i}}}(Q_{\bar{i}}T_{\bar{i}}(\lambda_{i})) & =& \textrm{Tr}_{V_{\bar{i}}}(Q_{\bar{i}}M_{\bar{i}}N_{\bar{i}}) \nonumber \\ & =& \textrm{Tr}_{V_{\bar{i}}\otimes V_{\ubar{i}}}(M_{\bar{i}}N_{\ubar{i}}^{t_{\ubar{i}}}\mathcal{Q}_{\bar{i}\ubar{i}})\,,
\end{eqnarray}
where the superscript $t_{\ubar{i}}$ in $N_{\ubar{i}}^{t_{\ubar{i}}}$ means the transpose of matrix $N$ in the new auxiliary space only. A graphical representation of this procedure is given in Fig.~\ref{fig:Tcartoon} following the minimalistic style common in the Bethe-ansatz literature (tensor-network style representations of the full construction will be given in Sec.~\ref{sec:conclusion}).
One can easily find that such a $\mathcal{Q}_{\bar{i}\ubar{i}}$ is given by
\begin{equation}\label{eq:Qmat2}
\mathcal{Q}_{\bar{i}\ubar{i}} = \tilde{Q}_{\bar{i}\ubar{i}} \cdot (Q_{\bar{i}} \otimes I_{\ubar{i}}) \,,
\end{equation} 
where
\begin{equation}
\tilde{Q}_{\bar{i}\ubar{i}} = 
\begin{pmatrix}1 & 0 & 0 & 1\\
0 & 0 & 0 & 0\\
0 & 0 & 0 & 0\\
1 & 0 & 0 & 1
\end{pmatrix}_{\bar{i}\ubar{i}}
\label{eq:tildeQ}
\end{equation} 
and the subscript of the right-hand side matrix means that it is written in the space $V_{\bar{i}}\otimes V_{\ubar{i}}$. 
Such a matrix appeared before in the ABA context for systems mixing particle and antiparticle representations \cite{Abad1996,Bolech2002,Bolech2005}.
Using this we have
\begin{multline}
\textrm{Tr}_{V_{\bar{i}}}(Q_{\bar{i}}T_{\bar{i}}(\lambda_{i})) = \textrm{Tr}_{V_{\bar{i}} \otimes V_{\ubar{i}}} [ L_{\bar{i}1}(\lambda_{i})\cdots L_{\bar{i}L}(\lambda_{i})\, \cdot \\
L_{\ubar{i}1}^{t_{\ubar{i}}}(\lambda_{i}) \cdots L_{\ubar{i}L}^{t_{\ubar{i}}}(\lambda_{i})  \cdot \mathcal{Q}_{\bar{i}\ubar{i}}]\,.
\end{multline} 
If we denote  
\begin{multline}
L_{i}(\lambda_{1},\cdots,\lambda_{n})=[L_{\bar{n}i}(\lambda_{n})L_{\ubar{n}i}^{t_{\ubar{n}}}(\lambda_{n})]\cdots \\ [L_{\bar{2}i}(\lambda_{2})L_{\ubar{2}i}^{t_{\ubar{2}}}(\lambda_{2})] [L_{\bar{1}i}(\lambda_{1})L_{\ubar{1}i}^{t_{\ubar{1}}}(\lambda_{1})]\,,
\end{multline}  
then Eq.~(\ref{eq:MPS1}) becomes
\begin{multline}\label{eq:MPS2}
\left|\lambda_{1},\lambda_{2},\cdots,\lambda_{n}\right\rangle =\textrm{Tr}_{(\bar{V} \otimes \ubar{V})^{\otimes n}}\{\left[L_{1}(\lambda_{1},\cdots,\lambda_{n})\left|\uparrow\right\rangle _{1}\right] \cdots \\ \left[L_{L}(\lambda_{1},\cdots,\lambda_{n})\left|\uparrow\right\rangle _{L}\right]\cdot(\mathcal{Q}_{\bar{n}\ubar{n}} \cdots \mathcal{Q}_{\bar{2}\ubar{2}}\mathcal{Q}_{\bar{1}\ubar{1}})\}\,,
\end{multline}
where $(\bar{V} \otimes \ubar{V})^{\otimes n}$ means the Hilbert space $V_{\bar{n}}\otimes V_{\ubar{n}}\otimes\dots\otimes V_{\bar{2}}\otimes V_{\ubar{2}}\otimes V_{\bar{1}}\otimes V_{\ubar{1}}$. 

It is convenient to introduce two matrices, $D_{n}$ and $C_{n}$, defined via
\begin{multline}  
L_{i}(\lambda_{1},\cdots,\lambda_{n})\left|\uparrow\right\rangle _{i}= \\  D_{n}(\lambda_{1},\cdots,\lambda_{n})\left|\uparrow\right\rangle _{i}+C_{n}(\lambda_{1},\cdots,\lambda_{n})\left|\downarrow\right\rangle_{i}\,, 
\end{multline}  
and one should keep in mind that $D_{n}$ and $C_{n}$ are $2^{2n} \times 2^{2n}$ matrices acting on the space $V_{\bar{n}}\otimes V_{\ubar{n}}\otimes\cdots\otimes V_{\bar{2}}\otimes V_{\ubar{2}}\otimes V_{\bar{1}}\otimes V_{\ubar{1}}$. Adopting the usual notations
\begin{equation}  
b(\lambda)\equiv\frac{\sinh(\lambda-\frac{\eta}{2})}{\sinh(\lambda+\frac{\eta}{2})}
\quad\text{and}\quad
c(\lambda)\equiv\frac{\sinh(\eta)}{\sinh(\lambda+\frac{\eta}{2})}\,,  
\end{equation} 
one can easily obtain
\begin{equation}
D_{1}(\lambda_{1})=\begin{pmatrix}1 & 0 & 0 & 0\\
0 & b(\lambda_{1}) & 0 & 0\\
0 & 0 & b(\lambda_{1}) & 0\\
c(\lambda_{1})^{2} & 0 & 0 & b(\lambda_{1})^{2}
\end{pmatrix}_{\bar{1}\ubar{1}}
\end{equation} 
and 
\begin{equation}
C_{1}(\lambda_{1})=\begin{pmatrix}0 & 0 & c(\lambda_{1}) & 0\\
b(\lambda_{1})c(\lambda_{1}) & 0 & 0 & b(\lambda_{1})c(\lambda_{1})\\
0 & 0 & 0 & 0\\
0 & 0 & c(\lambda_{1}) & 0
\end{pmatrix}_{\bar{1}\ubar{1}}\,.
\end{equation}
For $n>1$, recursion relations between $D_{n+1}$, $C_{n+1}$ and $D_{n}$, $C_{n}$ can be found as follows. Starting from the recursion
%\begin{multline}
$L_{i}(\lambda_{1},\cdots,\lambda_{n},\lambda_{n+1})=L_{\overline{(n+1)}i}(\lambda_{n+1})
\cdot L_{\underline{(n+1)}i}^{t_{\underline{(n+1)}}}(\lambda_{n+1})L_{i}(\lambda_{1},\cdots,\lambda_{n})$,
%\end{multline}
we can introduce $D_{n}$ and $C_{n}$, and derive expressions for $D_{n+1}$ and $C_{n+1}$. These are given by
\begin{multline}
D_{n+1}=
\begin{pmatrix}1 & 0 & 0 & 0\\
0 & b_{n+1} & 0 & 0\\
0 & 0 & b_{n+1} & 0\\
c_{n+1}^{2} & 0 & 0 & b_{n+1}^{2}
\end{pmatrix}_{\overline{(n+1)}\underline{(n+1)}}\otimes D_{n} +         \\
\begin{pmatrix}0 & c_{n+1} & 0 & 0\\
0 & 0 & 0 & 0\\
b_{n+1}c_{n+1} & 0 & 0 & b_{n+1}c_{n+1}\\
0 & c_{n+1} & 0 & 0
\end{pmatrix}_{\overline{(n+1)}\underline{(n+1)}}\otimes C_{n}\,,       
\end{multline}
and
\begin{multline}
C_{n+1}= 
\begin{pmatrix}b_{n+1}^{2} & 0 & 0 & c_{n+1}^{2}\\
0 & b_{n+1} & 0 & 0\\
0 & 0 & b_{n+1} & 0\\
0 & 0 & 0 & 1
\end{pmatrix}_{\overline{(n+1)}\underline{(n+1)}}\otimes C_{n} + \\
\begin{pmatrix}0 & 0 & c_{n+1} & 0\\
b_{n+1}c_{n+1} & 0 & 0 & b_{n+1}c_{n+1}\\
0 & 0 & 0 & 0\\
0 & 0 & c_{n+1} & 0
\end{pmatrix}_{\overline{(n+1)}\underline{(n+1)}}\otimes D_{n}\,,  
\end{multline}
where $b_{n+1}$, $c_{n+1}$ are shorthand notations for $b(\lambda_{n+1})$, and $c(\lambda_{n+1})$, respectively.

Using the above definitions one can, after a little algebra, achieve the goal of rewriting the Bethe state $\left|\lambda_{1},\lambda_{2},\cdots,\lambda_{n}\right\rangle $ as a MPS involving a trace over all the auxiliary spaces:
\begin{multline}\label{eq:ABA_MPS} 
\left|\lambda_{1},\lambda_{2},\cdots,\lambda_{n}\right\rangle =
\sum_{\{x_{1},x_{2},\cdots,x_{n}\}}
\mathrm{Tr}_{(\bar{V} \otimes \ubar{V})^{\otimes n}} 
\Big[ \left(D_{n}\right)^{x_{1}-1} \cdot  \\ 
C_{n} \left(D_{n}\right)^{x_{2}-x_{1}-1}C_{n}\cdots
\left(D_{n}\right)^{x_{n}-x_{n-1}-1}C_{n} \,\cdot \\
\left(D_{n}\right)^{L-x_{n}}\mathcal{Q}_{n} \Big] 
\cdot \left|x_{1},x_{2},\cdots,x_{n}\right\rangle\,, 
\end{multline}
where
\begin{equation} \label{eq:Qmat3}
\mathcal{Q}_{n}=\mathcal{Q}_{\bar{n}\ubar{n}}\otimes\cdots
\otimes\mathcal{Q}_{\bar{2}\ubar{2}}\otimes\mathcal{Q}_{\bar{1}\ubar{1}}\,,
\end{equation}
and $\left|x_{1},x_{2},\cdots,x_{n}\right\rangle$, with $1\leq x_{1}<x_{2}<\cdots<x_{n}\leq L$, denotes the configuration with down spins at those (lattice) locations; ---remember that $n$ is the total number of down spins. 

Remarkably, the MPSs obtained from ABA with OBC have the same form as the ones commonly postulated for the PBC case (ABA with PBC was giving the same structure, but it was as expected in that case~\cite{Katsura2010}). The correspondence is given by identifying the matrices $A_i^{\sigma_i=\uparrow}=D_n$ and $A_i^{\sigma_i=\downarrow}=C_n$, notice there is no dependence on the site index $i$, a kind of translational invariance. Of course, the appearance of the matrix $\mathcal{Q}_{n}$ is an additional feature that alters the translational invariance of the ansatz (for PBC, it was actually used to fix the lattice-momentum of the state~\cite{Alcaraz2004}). 
With OBC, the location of $\mathcal{Q}_{n}$ inside the trace indicates the location of the boundary, and one could also embed additional information on it to describe, for example, the presence of an impurity or a change of boundary conditions, (cf.~Ref.~\onlinecite{Draxler2017} for that kind of use but with PBC).
In Eq.~(\ref{eq:MPS2}), $\mathcal{Q}_{n}$ serves as a projector into the subspace with a fixed number of spin flips, $n$, but that has already been taken care of explicitly in the summation of Eq.~(\ref{eq:ABA_MPS}). In either case, $\mathcal{Q}_{n}$ has still the additional algebraic purpose of switching the horizontal boundary conditions of the ABA-generated six-vertex model into DWBC. The import of this is better illustrated by a specific example.

\subsection{A simple numerical example}

Let us consider the simplest non-trivial example to illustrate the way our MPS construction works. For the Hamiltonian in Eq.~(\ref{eq:HamXXZ}), we choose $L=3$ and only one down spin. The dimension of the Hilbert space is thus $3$. For such a small system, by direct calculation one can obtain the eigen\-energies, $E_{i}$, and the corresponding (non-normalized) eigenstates, $|\phi_{i} \rangle$, rather easily: 
\begin{enumerate}
\item $E_{1}=-3 \Delta$ with $|\phi_{1} \rangle = \left|\uparrow\right \rangle_{1}  \left| \uparrow \right\rangle_{2}  \left| \downarrow \right\rangle_{3}  -  \left| \downarrow \right\rangle_{1}  \left| \uparrow \right\rangle_{2}  \left| \uparrow \right\rangle_{3}$, 
\item $E_{2}=-4 \Delta + \sqrt{\Delta^{2}+8}$\quad with $| \phi_{2} \rangle = \left| \uparrow \right\rangle_{1}  \left| \uparrow \right\rangle_{2}  \left| \downarrow \right\rangle_{3}  +  \frac{1}{2} (\sqrt{\Delta^{2}+8}-\Delta) \left| \uparrow \right\rangle_{1}  \left| \downarrow \right\rangle_{2}  \left| \uparrow \right\rangle_{3}  +  \left| \downarrow \right\rangle_{1}  \left| \uparrow \right\rangle_{2}  \left| \uparrow \right\rangle_{3}$, 
\item $E_{3}=-4 \Delta - \sqrt{\Delta^{2}+8}$\quad with $| \phi_{3} \rangle = \left| \uparrow \right\rangle_{1}  \left| \uparrow \right\rangle_{2}  \left| \downarrow \right\rangle_{3}  -  \frac{1}{2} (\sqrt{\Delta^{2}+8}+\Delta) \left| \uparrow \right\rangle_{1}  \left| \downarrow \right\rangle_{2}  \left| \uparrow \right\rangle_{3}  +  \left| \downarrow \right\rangle_{1}  \left| \uparrow \right\rangle_{2}  \left| \uparrow \right\rangle_{3}$. 
\end{enumerate}
Next, we solve the same eigenvalue problem by the Bethe ansatz and use the MPS formulas we derived above. To be specific, let us choose $\Delta=2$, and then $\eta \approx 1.317$. Solving Eq.~(\ref{eq:BAE_XXZ}), we obtain three possible solutions $\lambda_{1}^{(1)}=i \pi/2$, or $\lambda_{1}^{(2)} \approx i 0.3747$, or $\lambda_{1}^{(3)} \approx -0.831 + i \pi/2$. Then it is very easy to check that Eq.~(\ref{eq:E_XXZ}) exactly recovers the three energies found by exact diagonalization. Further, one can use Eq.~(\ref{eq:ABA_MPS}) to get an expression for the corresponding eigenstates: $|\lambda_{1}^{(i)}\rangle = 
2 b^2(\lambda_{1}^{(i)})\,\left| \uparrow \right\rangle_{1}  \left| \uparrow \right\rangle_{2}  \left| \downarrow \right\rangle_{3}  
+ b(\lambda_{1}^{(i)}) \Big[1+b^2(\lambda_{1}^{(i)})+c^2(\lambda_{1}^{(i)})\Big]\,\left| \uparrow \right\rangle_{1}  \left| \downarrow \right\rangle_{2}  \left| \uparrow \right\rangle_{3}+  \Big[1+c^2(\lambda_{1}^{(i)})+b^2(\lambda_{1}^{(i)})c^2(\lambda_{1}^{(i)})+b^4(\lambda_{1}^{(i)})\Big]\,\left| \downarrow \right\rangle_{1}  \left| \uparrow \right\rangle_{2}  \left| \uparrow \right\rangle_{3}$; which, as can be seen by simple evaluation, are in agreement (up to normalization) with the $|\phi_{i}\rangle$ found directly. Of course, these eigenstates need not be and are not translationally invariant. One would expect that removing the matrix $\mathcal{Q}_n$ from the expression for the eigenstates will restore that invariance (and be unphysical), but the effect is actually more dramatic and they all become null vectors (i.e., invalid eigenstates). 

\begin{figure}
\begin{tikzpicture}[->,>=stealth',shorten >=1pt,auto,node distance=1.5cm,on grid,thick,
every state/.style={fill=yellow!30,draw=none,circular drop shadow,text=black}]
\node[state,fill=blue!30]            (L-00-L)     {$\left|\uparrow\right>_L$};
\node[state] (L-u1-L)  [above=of L-00-L]     {$L^{t_{\ubar{1}}}_{\ubar{1}L}$} edge [<-] (L-00-L);
\node[state] (L-b1-L)  [above=of L-u1-L]     {$L_{\bar{1}L}$}                            edge [<-] (L-u1-L);
\node[state] (L-u2-L)  [above=of L-b1-L]     {$L^{t_{\ubar{2}}}_{\ubar{2}L}$} edge [<-] (L-b1-L);
\node[state] (L-b2-L)  [above=of L-u2-L]     {$L_{\bar{2}L}$}                            edge [<-] (L-u2-L);
\node           (L-u3-L)  [above=of L-b2-L]     {$\vdots$}                                                         edge [-] (L-b2-L);
\node[state] (L-un-L)  [above=of L-u3-L]     {$L^{t_{\ubar{n}}}_{\ubar{n}L}$} edge [<-] (L-u3-L);
\node[state] (L-bn-L)  [above=of L-un-L]     {$L_{\bar{n}L}$}                            edge [<-] (L-un-L);
\node           (L-00-3)  [left=of L-00-L]         {$\cdots$};
\node           (L-u1-3)  [left=of L-u1-L]         {$\cdots$};
\node           (L-b1-3)  [above=of L-u1-3]     {$\cdots$};
\node           (L-u2-3)  [above=of L-b1-3]     {$\cdots$};
\node           (L-b2-3)  [above=of L-u2-3]     {$\cdots$};
\node           (L-u3-3)  [above=of L-b2-3]     {$\,$};
\node           (L-un-3)  [above=of L-u3-3]     {$\cdots$};
\node           (L-bn-3)  [above=of L-un-3]     {$\cdots$};
\node[state,fill=blue!30]           (L-00-2)  [left=of L-00-3] {$\left|\uparrow\right>_2$};
\node[state] (L-u1-2)  [left=of L-u1-3]         {$L^{t_{\ubar{1}}}_{\ubar{1}2}$} edge [<-] (L-00-2);
\node[state] (L-b1-2)  [above=of L-u1-2]     {$L_{\bar{1}2}$}                            edge [<-] (L-u1-2);
\node[state] (L-u2-2)  [above=of L-b1-2]     {$L^{t_{\ubar{2}}}_{\ubar{2}2}$} edge [<-] (L-b1-2);
\node[state] (L-b2-2)  [above=of L-u2-2]     {$L_{\bar{2}2}$}                            edge [<-] (L-u2-2);
\node           (L-u3-2)  [above=of L-b2-2]     {$\vdots$}                                                         edge [-] (L-b2-2);
\node[state] (L-un-2)  [above=of L-u3-2]     {$L^{t_{\ubar{n}}}_{\ubar{n}2}$} edge [<-] (L-u3-2);
\node[state] (L-bn-2)  [above=of L-un-2]     {$L_{\bar{n}2}$}                            edge [<-] (L-un-2);
\node[state,fill=blue!30]            (L-00-1)  [left=of L-00-2] {$\left|\uparrow\right>_1$};
\node[state] (L-u1-1)  [left=of L-u1-2]         {$L^{t_{\ubar{1}}}_{\ubar{1}1}$} edge [<-] (L-00-1);
\node[state] (L-b1-1)  [above=of L-u1-1]     {$L_{\bar{1}1}$}                            edge [<-] (L-u1-1);
\node[state] (L-u2-1)  [above=of L-b1-1]     {$L^{t_{\ubar{2}}}_{\ubar{2}1}$} edge [<-] (L-b1-1);
\node[state] (L-b2-1)  [above=of L-u2-1]     {$L_{\bar{2}1}$}                            edge [<-] (L-u2-1);
\node            (L-u3-1)  [above=of L-b2-1]     {$\vdots$}                                                        edge [-] (L-b2-1);
\node[state] (L-un-1)  [above=of L-u3-1]     {$L^{t_{\ubar{n}}}_{\ubar{n}1}$} edge [<-] (L-u3-1);
\node[state] (L-bn-1)  [above=of L-un-1]     {$L_{\bar{n}1}$}                            edge [<-] (L-un-1);
\node[state,fill=red!60] (Q-b1)  [left=of L-b1-1]        {$Q_{\bar{1}}$};
\node[state,fill=red!60] (Q-b2)  [left=of L-b2-1]        {$Q_{\bar{2}}$};
\node[state,fill=red!60] (Q-bn)  [left=of L-bn-1]        {$Q_{\bar{n}}$};
\path (L-u1-L) edge [-] (L-u1-3) (L-u1-3) edge (L-u1-2) (L-u1-2) edge (L-u1-1);
\path (L-b1-L) edge [-] (L-b1-3) (L-b1-3) edge (L-b1-2) (L-b1-2) edge (L-b1-1) (L-b1-1) edge (Q-b1);
\path (L-u2-L) edge [-] (L-u2-3) (L-u2-3) edge (L-u2-2) (L-u2-2) edge (L-u2-1);
\path (L-b2-L) edge [-] (L-b2-3) (L-b2-3) edge (L-b2-2) (L-b2-2) edge (L-b2-1) (L-b2-1) edge (Q-b2);
\path (L-un-L) edge [-] (L-un-3) (L-un-3) edge (L-un-2) (L-un-2) edge (L-un-1);
\path (L-bn-L) edge [-] (L-bn-3) (L-bn-3) edge (L-bn-2) (L-bn-2) edge (L-bn-1) (L-bn-1) edge (Q-bn);
\path (L-u1-L) edge [<->,in=0,out=0] (L-b1-L) (L-u1-1) edge [-,in=180,out=180] (Q-b1); 
\path (L-u2-L) edge [<->,in=0,out=0] (L-b2-L) (L-u2-1) edge [-,in=180,out=180] (Q-b2); 
\path (L-un-L) edge [<->,in=0,out=0] (L-bn-L) (L-un-1) edge [-,in=180,out=180] (Q-bn);
%\draw [blue] (current bounding box.south west) rectangle (current bounding box.north east);
\end{tikzpicture}
\caption{\label{fig:OBCnetwork} Schematic depiction of the tensor-network structure underlying the ABA construction of exact eigenstates for the six-vertex Heisenberg spin chain with OBC. The different nodes are labeled using the same notation as in the text. The vertical tensor contraction along each column of the network bracketed with $\left<\sigma_i\right|$ gives the matrices $A_i^{\sigma_i}$ that enter into the MPS rewriting of the eigenstates (i.e., $C_n$ and $D_n$). The horizontal dimension corresponds to the auxiliary space(s) of the ABA and also of the MPS construction ---the so-called \textit{bond dimension}. Those horizontal bonds are eventually all traced out, but in an unusual way due to the introduction of the matrix $\tilde{Q}$. The remaining piece of the $Q$-matrix sits on the left-most column, is built as a pure tensor product and has the combined functions of projector into the given number of excitations ($n$) and boundary-condition-changing operator imposing DWBC in the auxiliary space.}
\end{figure}
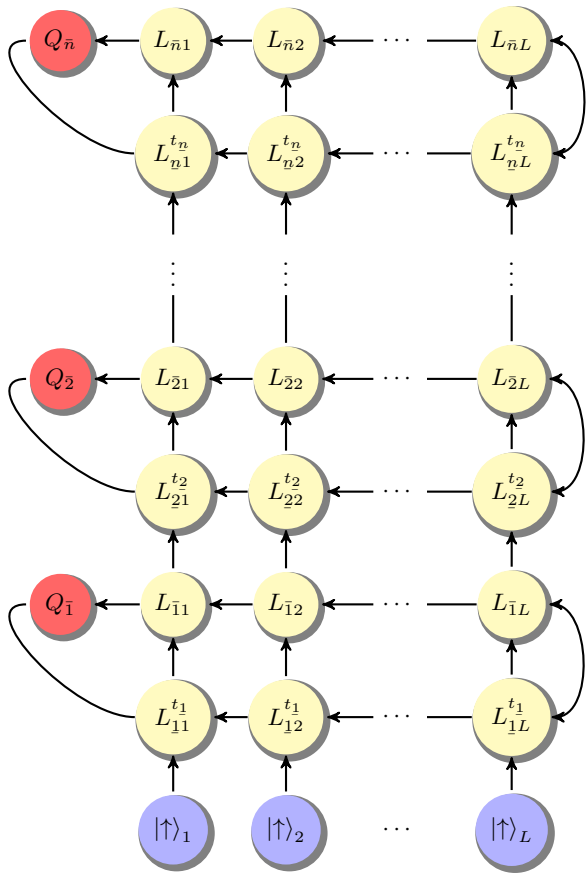

Repeating this exercise but for PBC, one finds translationally invariant eigenstates that are in agreement between exact diagonalization and the Bethe ansatz. Once again, removing the matrix $\mathcal{Q}_n$ from the ABA expression for the eigenstates will turn them all into null vectors. This suggests, as we discuss below, that any practical MPS ansatz should incorporate as an element the correlate of $\mathcal{Q}_n$.

\section{Discussion and Conclusions}\label{sec:conclusion}

\begin{figure}[b]
\begin{tikzpicture}[->,>=stealth',shorten >=1pt,auto,node distance=1.5cm,on grid,thick,
every state/.style={fill=yellow!30,draw=none,circular drop shadow,text=black}]
\path[use as bounding box] (-6.8, -0.5) rectangle (1, 6.8);
\node[state,fill=blue!30]            (L-00-L)     {$\left|\uparrow\right>_L$};
\node[state] (L-b1-L)  [above=of L-00-L]     {$L_{\bar{1}L}$}                            edge [<-] (L-00-L);
\node[state] (L-b2-L)  [above=of L-b1-L]     {$L_{\bar{2}L}$}                            edge [<-] (L-b1-L);
\node           (L-u3-L)  [above=of L-b2-L]     {$\vdots$}                                      edge [-] (L-b2-L);
\node[state] (L-bn-L)  [above=of L-u3-L]     {$L_{\bar{n}L}$}                            edge [<-] (L-u3-L);
\node           (L-00-3)  [left=of L-00-L]         {$\cdots$};
\node           (L-b1-3)  [left=of L-b1-L]     {$\cdots$};
\node           (L-b2-3)  [above=of L-b1-3]     {$\cdots$};
\node           (L-u3-3)  [above=of L-b2-3]     {$\,$};
\node           (L-bn-3)  [above=of L-u3-3]     {$\cdots$};
\node[state,fill=blue!30]           (L-00-2)  [left=of L-00-3] {$\left|\uparrow\right>_2$};
\node[state] (L-b1-2)  [left=of L-b1-3]     {$L_{\bar{1}2}$}                                edge [<-] (L-00-2);
\node[state] (L-b2-2)  [above=of L-b1-2]     {$L_{\bar{2}2}$}                            edge [<-] (L-b1-2);
\node           (L-u3-2)  [above=of L-b2-2]     {$\vdots$}                                       edge [-] (L-b2-2);
\node[state] (L-bn-2)  [above=of L-u3-2]     {$L_{\bar{n}2}$}                            edge [<-] (L-u3-2);
\node[state,fill=blue!30]            (L-00-1)  [left=of L-00-2] {$\left|\uparrow\right>_1$};
\node[state] (L-b1-1)  [left=of L-b1-2]     {$L_{\bar{1}1}$}                                edge [<-] (L-00-1);
\node[state] (L-b2-1)  [above=of L-b1-1]     {$L_{\bar{2}1}$}                            edge [<-] (L-b1-1);
\node            (L-u3-1)  [above=of L-b2-1]     {$\vdots$}                                      edge [-] (L-b2-1);
\node[state] (L-bn-1)  [above=of L-u3-1]     {$L_{\bar{n}1}$}                            edge [<-] (L-u3-1);
\node[state,fill=red!60] (Q-b1)  [left=of L-b1-1]        {$Q_{\bar{1}}$};
\node[state,fill=red!60] (Q-b2)  [left=of L-b2-1]        {$Q_{\bar{2}}$};
\node[state,fill=red!60] (Q-bn)  [left=of L-bn-1]        {$Q_{\bar{n}}$};
\path (L-b1-L) edge [-] (L-b1-3) (L-b1-3) edge (L-b1-2) (L-b1-2) edge (L-b1-1) (L-b1-1) edge (Q-b1);
\path (L-b2-L) edge [-] (L-b2-3) (L-b2-3) edge (L-b2-2) (L-b2-2) edge (L-b2-1) (L-b2-1) edge (Q-b2);
\path (L-bn-L) edge [-] (L-bn-3) (L-bn-3) edge (L-bn-2) (L-bn-2) edge (L-bn-1) (L-bn-1) edge (Q-bn);
\path (Q-b1) edge  [in=20,out=160,looseness=0.8] (L-b1-L);
\path (Q-b2) edge  [in=20,out=160,looseness=0.8] (L-b2-L);
\path (Q-bn) edge  [in=20,out=160,looseness=0.8] (L-bn-L);
%\draw [blue] (current bounding box.south west) rectangle (current bounding box.north east);
\end{tikzpicture}
\caption{\label{fig:PBCnetwork} Another schematic depiction of a tensor-network structure for  the six-vertex Heisenberg spin chain but this time for PBC. The different nodes are labeled, \textit{mutatis mutandis}, using a similar notation as in the text. The ABA construction is simpler in this case and the auxiliary or bond dimension is traced out in the conventional way. Notice one still finds a $Q$-matrix, the leftmost column in the network, that fixes the excitation number and modifies the horizontal boundary conditions of the ABA six-vertex construct.}
\end{figure}
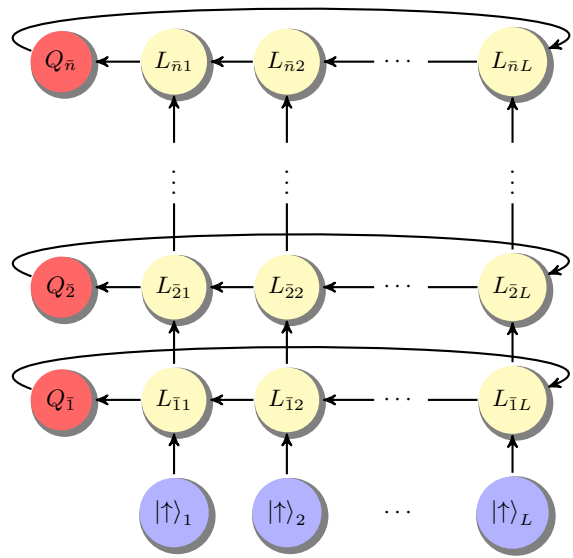

As already mentioned, looking at Eq.~(\ref{eq:ABA_MPS}) one finds that the MPS with OBC that we derived here is different from the one discussed in the current tensor-network literature, and it is closer instead to the one discussed for PBC~\cite{Schollwock2011}. The ABA-generated tensor-network structure that we found for OBC is illustrated in Fig.~\ref{fig:OBCnetwork}.  This has a more complicated but essentially similar global structure to that found for the PBC case. The latter was already derived elsewhere by an explicit ABA construction~\cite{Katsura2010}, and it is provided for comparison in Fig.~\ref{fig:PBCnetwork}.

The natural tensor-network construction emanating from ABA~\cite{MurgKorepin2012} does not fit either of the standard MPS ans\"atze for open or periodic boundaries that we reviewed in Sec.~\ref{sec:introduction}. Going beyond finite-system-size calculations and motivated by the infinite-system-size limit, one knows that the bulk physics does not depend on the boundary conditions and one should be able to reach the same so-called infinite-MPS (iMPS) or infinite-DMRG (iDMRG)~\cite{McCulloch2008,Kjall2013} construction starting from either; ---though the original DMRG work by White already explored the infinite-size limit, these ideas were first put to use in tensor-network language exploiting translational invariance in the work on (infinite) time evolving block decimation ((i)TEBD)~\cite{Vidal2007,OrusVidal2008}. Focusing on the common global structure of our OBC and the PBC Bethe states after casting them into traced matrix-product forms, we propose the following unified practical MPS ansatz:
\begin{equation}
\left|\mathrm{MPS}\right>=\sum_{\{\sigma\}}\mathrm{Tr}
\left[A_1^{\sigma_1} \cdots A_L^{\sigma_L}\mathcal{Q}^r\right]
\,\left|\sigma_1\dots\sigma_L\right>\,,
\end{equation}
where the $A_i^{\sigma_i}$ are $D \times D$ variational matrices. Notice that from the ABA we find $A_i^{\sigma}=A_j^{\sigma}\;\forall i,j$ and independently of the size of the system. We thus have the usual site independence of the $A$-matrices in the spirit of iMPS but without needing to invoke translational invariance. In contradistinction with the practice in the literature, ABA prompts the inclusion of an additional matrix into the ansatz that is also taken as a variational parameter (for a total of $d+1$ variational matrices). We introduce the integer index $r=0,1$ which we call the rank index of the ansatz, and we consider it also a variational parameter. Rank zero is equivalent to not having $\mathcal{Q}$ and falls back into the standard iMPS structure. Rank one can be distinctively different and we posit it as the generic choice for Bethe-ansatz-integrable systems.

While the $A$-matrices are taken as unrestricted, we expect from the ABA-based derivations that $\mathcal{Q}$ is nilpotent, i.e. $\mathcal{Q}^2=0$ ---though it could be generalized to higher degrees of nil\-potency. This can be achieved by the introduction of a generic Jordan form~\cite{Chung2015}, which can serve the dual purpose of fixing an arbitrary gauge-like degree of freedom typical of this kind of traced matrix products. Another alternative way of handling the nil\-potency constraint is motivated by a parametrization used for MPS-DMRG with OBC in which boundary vectors are introduced into the ansatz for the state components: $\left<\alpha\right|A^{\sigma_1}_1\cdots A^{\sigma_L}_L\left|\beta\right> = \mathrm{Tr}\left[A^{\sigma_1}_1\cdots A^{\sigma_L}_L \cdot
(\left|\beta\right>\left<\alpha\right|) \right]$. This parametrization can be interpreted in the framework of Fig.~\ref{fig:MPSwOBC} by taking all the telescoping matrices at either end of the chain and considering their product as defining the boundary vectors (which could be found, for instance, using exact diagonalization, cf.~\cite{Kancharla2001}; this point of view also resembles, but is different in spirit from, the idea of \textit{infinite boundary conditions} for the  iMPS ansatz~\cite{Phien2012}). Pictorially,
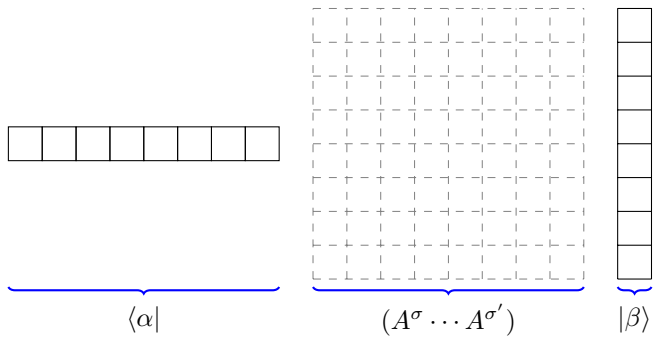
\begin{center}
\begin{tikzpicture}[scale=0.45, set style={{help lines}+=[dashed]}]
\foreach \x in {8,...,15} \draw (\x,3.5) rectangle +(1,1);
\draw[style=help lines] (17,0) grid +(8,8);
\draw (26,0) grid +(1,8);
\draw[snake=brace, raise snake=3pt, thick, blue] (16,0) -- +(-8,0);
\draw[snake=brace, raise snake=3pt, thick, blue] (25,0) -- +(-8,0);
\draw[snake=brace, raise snake=3pt, thick, blue] (27,0) -- +(-1,0);
\node at (12,-1.2) {$\left<\alpha\right|$};
\node at (21,-1.2) {$(A^{\sigma} \cdots A^{\sigma'})$};
\node at (26.5,-1.2) {$\left|\beta\right>$};
\end{tikzpicture}
\end{center}
so that $\mathcal{Q}\equiv\left|\beta\right>\left<\alpha\right|$ is given by the outer product of two variational vectors, and its nil\-potency is equivalent to requesting those vectors to be orthogonal, ---which is the case in the ABA constructions and in practice can be easily achieved by a Gram-Schmidt procedure.

Some forward-looking considerations might help put our results in perspective and provide some open-ended prospects for future research. In the case of the six-vertex spin chain, we expect due to the ABA construction that a rank-1 iMPS ansatz would be the better choice (in the large-$L$ limit and at constant bond dimension, otherwise a rank-0 ansatz can be rewritten as a rank-zero one by increasing $D\to LD$ and introducing an appropriate block structure). A natural model to numerically test the differences between the two ranks could be, for example, the bilinear-bi\-quadratic spin-1 chain or its extensions~\cite{Tsai2000,Lauchli2006,Andres2008,Pixley2014}. Its AKLT point~\cite{Affleck1987} in the interior of the Haldane phase would most likely be better described by a rank-zero ansatz; while its integrable point, the Tahkhtajan-Babudjian point~\cite{Takhtajan1982,Babujian1982}, at the interface between the Haldane and the dimer phases would likely call for a rank-one ansatz, (and similarly for the Lai-Sutherland point~\cite{Lai1974,Sutherland1975}, on the other end of the Haldane phase at the interface with the trimer phase, which has $\mathrm{SU}(3)$ symmetry and can be solved via a nested ABA). The rank index that we introduced might vary with the gapped versus gapless, or even integrable or topological character of the different phases, a hypothesis that can be first tested numerically and points to an interesting direction for further studies.\footnote{There are known properties that differentiate integrable from non-integrable systems. Like the statistics of energy level spacings \cite{Poilblanc1993} or certain related transport properties \cite{Narozhny1998}. As in these two examples, it could be that different-rank matrix-product ans\"atze will only differ when a finite-size regularization is applied, or that the difference will also be present  for infinite-size systems, (cf.~with our earlier discussion of DWBC).} 
Another ground for further exploration would be in the context of continuum models. Using the same ideas as above, we can obtain a MPS for a lattice-regularized version of an interacting gas model with OBC; the example of a single-component Bose gas is given in the appendix. 

\bigskip
\textit{In conclusion}, we have derived the MPS representation of the Bethe states for the six-vertex Heisenberg spin chain with OBC. 
We find important similarities and differences with the various MPSs discussed in the tensor-networks literature for both open and periodic boundary conditions.
In particular, our results clarify the conditions for the implementation of iMPS and prove that such an approach can address directly the thermodynamic limit in a unified way for both PBC and OBC. 
Our construction is rather generic and can be applied to other integrable models solvable via the standard ABA.
Moreover, following the general construction by Sklyanin \cite{Sklyanin1988} or other more recent developments \cite{Wang2015}, one can have also boundary scattering matrices factored into the looped monodromy matrix of the system and thus incorporate the effects of more general boundaries that include, for example, the effects of arbitrarily oriented boundary fields.
This technical progress goes hand in hand with new physical insights.
The unified ansatz for OBC/PBC reveals how to apply the physically motivated notion of (local) symmetry under translations regardless of the boundary conditions,
thus indicating the proper way of considering the infinite-size limit within an MPS-type ansatz, like iMPS, but in a way that is, remarkably, also valid for finite-size systems and can more naturally be combined with finite-size scaling.
In doing so, our results also justify how to incorporate boundary degrees of freedom and how their influence on the bulk can be computed (effects such as Friedel oscillations, etc.).
Since the MPS matrices can be taken to be always the same, in particular at the boundary sites where we have additional matrices encoding the boundary effects, the whole physics is contained in the boundary-MPS degrees of freedom, thus supporting the viability of the ideas of boundary-bulk correspondence for these types of systems.
In retrospect, DMRG and related algorithms were originally motivated by considerations emanating from statistical mechanics and renormalization-group theory. 
Later, ideas from quantum information theory were instrumental in further developing the algorithms. 
Ours and related work bring in additional insights from the theory of integrable systems that shed new light and can help advance the algorithms even further. 
Our work makes the Bethe eigenstates more accessible and thus informs the choice of ansatz for numerical solutions of both integrable and non-integrable systems in one dimension.

\begin{acknowledgments}
We are grateful for the hospitality of the Kavli Institute for Theoretical Physics at UC Santa Barbara (funded under NSF PHY11-25915) where we had interesting discussions with J.~I.~Cirac, J.~Haegeman and F.~Verstraete, and also the last stages of the writing took place. 
\end{acknowledgments}

\appendix

\section{Matrix product states for the Lieb-Liniger model with OBC}\label{sec:ABA of Lieb-Liniger}

The Hamiltonian of the Lieb-Liniger model is given by
\begin{equation}
H=\int_{0}^{L}[\partial_{x}\psi^{\dagger}(x)\partial_{x}\psi(x)+\kappa\psi^{\dagger}(x)\psi^{\dagger}(x)\psi(x)\psi(x)]dx \nonumber,
\end{equation}
where $\kappa>0$ and we have fixed $\hbar=1=2m$. The bosonic field operators satisfy the canonical commutation relation $[\psi(x),\psi^{\dagger}(y)]=\delta(x-y)$. For the $n$-particle state $\left|\varPsi_{0}\right\rangle =\int dx_{1}\cdots dx_{n}\varPsi_{0}(x_{1},\ldots,x_{n})\psi^{\dagger}(x_{1})\cdots\psi^{\dagger}(x_{n})\left|0\right\rangle $, one can derive the Schr\"odinger equation from the above Hamiltonian~\cite{LiebLiniger1963}:
\begin{equation}
[-\sum_{j=1}^{n}\partial_{x_{j}}^{2}+2\kappa\sum_{1\leq j<j^{'}\leq n}\delta(x_{j}-x_{j^{'}})]\varPsi_{0}=E\varPsi_{0},
\end{equation}
where $\varPsi_{0}=\varPsi_{0}(x_{1},\ldots,x_{n})$. We shall now introduce a lattice-regularized version of this model~\cite{Tarasov1983} (which is more general than studying the dilute limit of Hubbard-type models which are not integrable in the bosonic case, cf.~\cite{Mei2016,Bolech2012}), where the spatial position $x\in \mathbb{R}$ in the continuum model is replaced by the site $i\in \mathbb{Z}$ with the lattice spacing $a$. Let $V_{i}$ be a physical Hilbert space at the $i$th site spanned by $\left|m\right\rangle _{i}=(1/\sqrt{m!})(\psi_{i}^{\dagger})^{m}\left|0\right\rangle _{i}$ with $m\geq0$. Here, $\psi_{i}^{\dagger}$ and $\psi_{i}$ are the bosonic creation and annihilation operators on $V_{i}$, respectively. They satisfy $[\psi_{i},\psi_{i^{'}}^{\dagger}]=\delta_{ii^{'}}$, and $[\psi_{i},\psi_{i^{'}}]=0$. In the continuum limit $(a\rightarrow0)$, $\psi_{i}\rightarrow\psi(x_{i})\sqrt{a}$. Note that $\psi_{i}$ and $\kappa a$ are dimensionless. There are $N$ sites in total, and $L=Na$.
 
The $L$-operator at the $i$th site is given by
\begin{multline*} 
L_{0i}(\lambda)=
\begin{pmatrix}1-\frac{i\lambda a}{2}+\frac{\kappa a}{2}\psi_{i}^{\dagger}\psi_{i} & -i\sqrt{\kappa a}\psi_{i}^{\dagger}\rho_{i}\\
i\sqrt{\kappa a}\rho_{i}\psi_{i} & 1+\frac{i\lambda a}{2}+\frac{\kappa a}{2}\psi_{i}^{\dagger}\psi_{i}
\end{pmatrix}_{0}\,, 
\end{multline*} 
where
%\begin{eqnarray} 
$\rho_{i}=\sqrt{1+\frac{\kappa a}{4}\psi_{i}^{\dagger}\psi_{i}}$.
%\end{eqnarray}  
The matrix elements of $L_{0i}$ and $\rho_{i}$ are operators acting on the space $V_{i}$ and $L_{0i}$ is represented as a $2 \times 2$ matrix in the auxiliary space $V_{0}$. This $L$-operator satisfies the following relation:
\begin{equation}\label{eq:YBE_LL}
R_{ab}(\lambda-\mu)L_{ai}(\lambda)L_{bi}(\mu)=
L_{bi}(\mu)L_{ai}(\lambda)R_{ab}(\lambda-\mu)\,,  
\end{equation} 
with the $R$-matrix given by
 \begin{equation}
 R_{ab}(\lambda)=\begin{pmatrix}1 & 0 & 0 & 0\\
0 & \frac{\lambda}{\lambda-i\kappa} & -\frac{i\kappa}{\lambda-i\kappa} & 0\\
0 & -\frac{i\kappa}{\lambda-i\kappa} & \frac{\lambda}{\lambda-i\kappa} & 0\\
0 & 0 & 0 & 1
\end{pmatrix}_{ab}\,.
 \end{equation}

To construct the looped monodromy matrix of the Lieb-Liniger model with OBC, we need to know the inverse of the $L$-operator. One can write $L_{ai}(\lambda)$ as
\begin{equation}
 L_{ai}(\lambda)=\begin{pmatrix}
\mathcal{A}(\lambda) & \mathcal{B}(\lambda) \\
\mathcal{C}(\lambda) & \mathcal{D}(\lambda) 
\end{pmatrix}_{a}\,.
\end{equation}
Choosing $\lambda - \mu = i \kappa$ in Eq.~(\ref{eq:YBE_LL}), one obtains the following equations
\begin{align*}
\mathcal{A}(\mu - \frac{i \kappa}{2}) \mathcal{D}(\mu + \frac{i \kappa}{2}) - \mathcal{B}(\mu - \frac{i \kappa}{2}) \mathcal{C}(\mu + \frac{i \kappa}{2}) =&   \\
\textrm{det}_{q}& L_{0i}(\mu)\,,        \\
\mathcal{B}(\mu - \frac{i \kappa}{2}) \mathcal{A}(\mu + \frac{i \kappa}{2}) - \mathcal{A}(\mu - \frac{i \kappa}{2}) \mathcal{B}(\mu + \frac{i \kappa}{2}) =& \;0\,,        \\
\mathcal{C}(\mu - \frac{i \kappa}{2}) \mathcal{D}(\mu + \frac{i \kappa}{2}) - \mathcal{D}(\mu - \frac{i \kappa}{2}) \mathcal{C}(\mu + \frac{i \kappa}{2}) =& \;0\,,       \\
\mathcal{D}(\mu - \frac{i \kappa}{2}) \mathcal{A}(\mu + \frac{i \kappa}{2}) - \mathcal{C}(\mu - \frac{i \kappa}{2}) \mathcal{B}(\mu + \frac{i \kappa}{2}) =&   \\
\textrm{det}_{q}& L_{0i}(\mu)\,,        
\end{align*} 
where $\textrm{det}_{q} L_{0i}(\mu)$ is known as the quantum determinant~\cite{Korepin1993} and is given by
%\begin{eqnarray}
$\textrm{det}_{q} L_{0i}(\mu) = \frac{a^{2}}{4}(\mu - \frac{i \kappa}{2} + \frac{2 i}{a}) (\mu + \frac{i \kappa}{2} - \frac{2 i}{a})$.
%\end{eqnarray} 
Using these four relations, we obtain
\begin{equation}
L_{0i}(\mu - \frac{i \kappa}{2}) \sigma_{0}^{y} L_{0i}^{t_{0}}(\mu + \frac{i \kappa}{2}) \sigma_{0}^{y} = \left[ \textrm{det}_{q} L_{0i}(\mu) \right] \cdot I_{0}\,,  
\end{equation}
where $I_{0}$ is the identity matrix in the auxiliary space $V_{0}$. Then, it follows immediately that
\begin{equation*}
L_{0i}^{-1}(\mu)=\frac{4}{a^{2}(\mu+\frac{2i}{a})(\mu+i\kappa-\frac{2i}{a})}\sigma_{0}^{y}L_{0i}^{t_{0}}(\mu+i\kappa)\sigma_{0}^{y}\,.
\end{equation*}
Then, the looped monodromy matrix of the Lieb-Liniger model with OBC is given by (after shifting the parameter and omitting a constant factor) 
\begin{multline}
T_{0}(\lambda) = 
L_{01}(\lambda+\frac{i\kappa}{2}) \cdots L_{0N}(\lambda+\frac{i\kappa}{2}) \cdot \\
\Big[ \sigma_{0}^{y} L_{0N}^{t_{0}}(-\lambda + \frac{i\kappa}{2}) \sigma_{0}^{y}\Big] \cdots \Big[\sigma_{0}^{y}L_{01}^{t_{0}}(-\lambda+\frac{i\kappa}{2}) \sigma_{0}^{y}\Big]\,.
\end{multline}

The loop monodromy matrix satisfies the fundamental commutation relation:
\begin{multline}
R_{00^{'}}(\lambda-\mu)T_{0}(\lambda)R_{00^{'}}(\lambda+\mu+i\kappa)T_{^{0'}}(\mu) = \\
T_{^{0'}}(\mu)R_{00^{'}}(\lambda+\mu+i\kappa)T_{0}(\lambda)R_{00^{'}}(\lambda-\mu)\,.
\end{multline}
In the auxiliary space $V_{0}$, the matrix can be written in block form as
\begin{equation}
T_{0}(\lambda)=\begin{pmatrix}A(\lambda) & B(\lambda)\\
C(\lambda) & D(\lambda)
\end{pmatrix}_{0}
\end{equation}
and the transfer matrix is $t(\lambda) = \textrm{tr}_{V_{0}}T_{0}(\lambda) = A(\lambda) + D(\lambda)$. One can easily verify that $t(\lambda)t(\mu)=t(\mu)t(\lambda)$~\cite{Sklyanin1988}. 
A Bethe eigenstate of the Hamiltonian is then given by
\begin{equation}
\left|\lambda_{1},\lambda_{2},\cdots,\lambda_{n}\right\rangle =B(\lambda_{n})\cdots B(\lambda_{2})B(\lambda_{1})\left|0\right\rangle ^{(N)}\,,
\end{equation}
where $\left|0\right\rangle ^{(N)}=\left|0\right\rangle _{1}\left|0\right\rangle _{2}\cdots\left|0\right\rangle _{N}$ is the global vacuum state and the $| 0 \rangle _{i}$ are the local vacuum states at sites labeled by $i$. From the fundamental commutation relation, we know that $B(\lambda)B(\mu)=B(\mu)B(\lambda)$. Using the general procedure of the quantum inverse scattering method for OBC~\cite{Sklyanin1988}, one gets the following Bethe ansatz equations:
\begin{multline}
e^{i2\lambda_{i}L} = \prod_{\substack{j=1\\ j\neq i}}^{n}
\frac{(\lambda_{i}+\lambda_{j}+i\kappa)(\lambda_{i}-\lambda_{j}+i\kappa)}{(\lambda_{i}+\lambda_{j}-i\kappa)(\lambda_{i}-\lambda_{j}-i\kappa)}\,,  \\
i = 1, 2, ..., n\,,
\end{multline}
which are consistent with the results in~\cite{Gaudin1971}.

\bigskip

We move on to derive a MPS for the lattice version of the Lieb-Liniger model with OBC. Using
%\begin{eqnarray}
$B(\lambda)= \hspace{0.001 in} _{0}\!\left\langle \uparrow\right|T_{0}(\lambda)
\left|\downarrow\right\rangle _{0}$,
%\end{eqnarray}
one obtains
\begin{multline}
\left|\lambda_{1},\lambda_{2},\cdots,\lambda_{n}\right\rangle =[\textrm{Tr}_{V_{\bar{n}}}(Q_{\bar{n}}T_{\bar{n}}(\lambda_{n}))]\cdots \\
[\textrm{Tr}_{V_{\bar{2}}}(Q_{\bar{2}}T_{\bar{2}}(\lambda_{2}))][\textrm{Tr}_{V_{\bar{1}}}(Q_{\bar{1}}T_{\bar{1}}(\lambda_{1}))]\cdot\left|0\right\rangle ^{(N)}\,,
\end{multline}
where the matrix $Q_{\bar{i}}$ is the same as the one in Eq.~(\ref{eq:Qmat}). 
Defining
%\begin{eqnarray}
$A_{\bar{n}i}(\lambda_{n})=L_{\bar{n}i}(\lambda_{n}+\frac{i\kappa}{2})$ and
$B_{\bar{n}i}(\lambda_{n})=\sigma_{\bar{n}}^{y}L_{\bar{n}i}^{t_{\bar{n}}}(-\lambda_{n}+\frac{i\kappa}{2})\sigma_{\bar{n}}^{y}$,
%\end{eqnarray}
then
%\begin{eqnarray}
$\textrm{Tr}_{V_{\bar{n}}}[Q_{\bar{n}}T_{\bar{n}}(\lambda_{n})] = \textrm{Tr}_{V_{\bar{n}}}\{Q_{\bar{n}}\cdot [A_{\bar{n}1}(\lambda_{n})
\cdots A_{\bar{n}N}(\lambda_{n})]\cdot[B_{\bar{n}N}(\lambda_{n})\cdots B_{\bar{n}1}(\lambda_{n})]\}$.
%\end{eqnarray}
Using now the same method as in Sec.~\ref{sec:MPS for XXZ}, we obtain
\begin{multline}
\textrm{Tr}_{V_{\bar{n}}}[Q_{\bar{n}}T_{\bar{n}}(\lambda_{n})] = \textrm{Tr}_{V_{\bar{n}}\otimes V_{\ubar{n}}}\{[A_{\bar{n}1}(\lambda_{n})\cdots A_{\bar{n}N}(\lambda_{n})] \, \cdot \\
[B_{\ubar{n}1}^{t_{\ubar{n}}}(\lambda_{n})\cdots B_{\ubar{n}N}^{t_{\ubar{n}}}(\lambda_{n})]\cdot\mathcal{Q}_{\bar{n}\ubar{n}}\}\,,
\end{multline}
where $\mathcal{Q}_{\bar{n}\ubar{n}}$ is the same as in Eq.~(\ref{eq:Qmat2}), and
%\begin{eqnarray}
$B_{\ubar{n}i}^{t_{\ubar{n}}}(\lambda_{n})=\sigma_{\ubar{n}}^{y}L_{\ubar{n}i}(-\lambda_{n}+\frac{i\kappa}{2})\sigma_{\ubar{n}}^{y}$.
%\end{eqnarray}
With this, one can easily obtain
\begin{multline}
\left|\lambda_{1},\lambda_{2},\cdots,\lambda_{n}\right\rangle = 
\mathrm{Tr}_{(\bar{V} \otimes \ubar{V})^{\otimes n}}\Big[
L_{1}(\lambda_{1},\cdots,\lambda_{n})\left|0\right\rangle _{1}  \cdots \\
L_{N}(\lambda_{1},\cdots,\lambda_{n})\left|0\right\rangle _{N} \cdot
(\mathcal{Q}_{\bar{n}\ubar{n}}\cdots
\mathcal{Q}_{\bar{2}\ubar{2}}
\mathcal{Q}_{\bar{1}\ubar{1}}) \Big]\,,
\end{multline}
where $(\bar{V} \otimes \ubar{V})^{\otimes n}$ means $V_{\bar{n}}\otimes V_{\ubar{n}}\otimes\cdots\otimes V_{\bar{2}}\otimes V_{\ubar{2}}\otimes V_{\bar{1}}\otimes V_{\ubar{1}}$ as defined before, and here
%\begin{eqnarray}
$L_{i}(\lambda_{1},\cdots,\lambda_{n})=A_{\bar{n}i}(\lambda_{n})B_{\ubar{n}i}^{t_{\ubar{n}}}(\lambda_{n}) 
\cdots A_{\bar{2}i}(\lambda_{2})B_{\ubar{2}i}^{t_{\ubar{2}}}(\lambda_{2})A_{\bar{1}i}(\lambda_{1})B_{\ubar{1}i}^{t_{\ubar{1}}}(\lambda_{1})$.
%\end{eqnarray}

Let us introduce the notations
\begin{align}
\delta_{m}=&\;
i\sqrt{(m+1)\cdot\kappa a\cdot(1+\frac{m\kappa a}{4})}\,,   \\
\gamma_{n,m}=&\;
1-\frac{i(\lambda_{n}+\frac{i\kappa}{2})a}{2}+\frac{m\kappa a}{2}\,,   \\
\beta_{n,m}=&\;
1+\frac{i(\lambda_{n}+\frac{i\kappa}{2})a}{2}+\frac{m\kappa a}{2}\,,   
\end{align}
and also define
\begin{equation*}
C_{n,m}(\lambda_{1},\lambda_{2},\cdots,\lambda_{n})
=\,_{i}\!\left\langle m\right|L_{i}(\lambda_{1},\lambda_{2},\cdots,\lambda_{n})\left|0\right\rangle _{i}\,.
\end{equation*}
For a given $n$, it is obvious that $C_{n,m}\neq0$ only when $m=0,1,2,\cdots,2n$. Going forward, one should also keep in mind that $C_{n,m}$ are $2^{2n} \times 2^{2n}$ matrices acting on the space $(\bar{V} \otimes \ubar{V})^{\otimes n}$. We can easily obtain, for $n=1$,
%\begin{allowdisplaybreaks}
\begin{align*}
C_{1,0}(\lambda_{1})&=
\begin{pmatrix}\gamma_{1,0}\beta_{1,0}^{*} & 0 & 0 & 0\\
0 & \gamma_{1,0}\gamma_{1,0}^{*} & 0 & 0\\
0 & 0 & \beta_{1,0}\beta_{1,0}^{*} & 0\\
(\delta_{0})^{2} & 0 & 0 & \beta_{1,0}\gamma_{1,0}^{*}     
\end{pmatrix}_{\bar{1}\ubar{1}}\,,\\
%\end{equation*}
%\begin{equation*}                                                  
C_{1,1}(\lambda_{1})&=
\begin{pmatrix}0 & 0 & -\delta_{0}\beta_{1,0}^{*} & 0\\
\delta_{0}\gamma_{1,1} & 0 & 0 & -\delta_{0}\gamma_{1,0}^{*}\\
0 & 0 & 0 & 0\\
0 & 0 & \delta_{0}\beta_{1,1} & 0
\end{pmatrix}_{\bar{1}\ubar{1}}\quad\text{and} \\        
%\end{equation*}  
%\begin{equation*}                                          
C_{1,2}(\lambda_{1})&=
\begin{pmatrix}0 & 0 & 0 & 0\\
0 & 0 & -\delta_{0}\delta_{1} & 0\\
0 & 0 & 0 & 0\\
0 & 0 & 0 & 0
\end{pmatrix}_{\bar{1}\ubar{1}}\,.      
\end{align*}
%\end{allowdisplaybreaks}
(Here we have used the fact that all solutions of the Bethe-ansatz equations for the repulsive Lieb-Liniger model with OBC are real.) 
Using
\begin{multline}
C_{n+1,m}=~_{i}\!\left\langle m\right|
A_{\overline{(n+1)}i}(\lambda_{n+1}) \,\cdot\\
B_{\underline{(n+1)}i}^{t_{\underline{(n+1)}}}(\lambda_{n+1})
L_{i}(\lambda_{1},\cdots,\lambda_{n})
\left|0\right\rangle _{i}\,,
\end{multline} 
we can easily obtain with a little bit of algebra the recursion relation between $C_{n+1,m}$ and $C_{n,m^{'}}$, which is given by
\onecolumngrid                                                                                                                                       %One column
%\begin{widetext}
%\begin{allowdisplaybreaks}
\begin{align*}
C_{n+1,m}= &                                            
\begin{pmatrix}0 & 0 & 0 & 0\\
0 & 0 & 0 & 0\\
0 & -\delta_{m}\delta_{m+1} & 0 & 0\\
0 & 0 & 0 & 0
\end{pmatrix}_{\overline{(n+1)}\underline{(n+1)}}\otimes C_{n,m+2} \quad+\quad 
\begin{pmatrix}0 & 0 & 0 & 0\\
0 & 0 & -\delta_{m-1}\delta_{m-2} & 0\\
0 & 0 & 0 & 0\\
0 & 0 & 0 & 0
\end{pmatrix}_{\overline{(n+1)}\underline{(n+1)}}\otimes C_{n,m-2}\; + \\
&\begin{pmatrix}0 & -\delta_{m}\gamma_{n+1,m} & 0 & 0\\
0 & 0 & 0 & 0\\
\delta_{m}\beta_{n+1,m+1}^{*} & 0 & 0 & -\delta_{m}\beta_{n+1,m}\\
0 & \delta_{m}\gamma_{n+1,m+1}^{*} & 0 & 0
\end{pmatrix}_{\overline{(n+1)}\underline{(n+1)}}\otimes C_{n,m+1}\;  +   \\
&\begin{pmatrix}0 & 0 & -\delta_{m-1}\beta_{n+1,m-1}^{*} & 0\\
\delta_{m-1}\gamma_{n+1,m} & 0 & 0 & -\delta_{m-1}\gamma_{n+1,m-1}^{*}\\
0 & 0 & 0 & 0\\
0 & 0 & \delta_{m-1}\beta_{n+1,m} & 0
\end{pmatrix}_{\overline{(n+1)}\underline{(n+1)}}\otimes C_{n,m-1}\;   +    \\
&\begin{pmatrix}\gamma_{n+1,m}\beta_{n+1,m}^{*} & 0 & 0 & (\delta_{m-1})^{2}\\
0 & \gamma_{n+1,m}\gamma_{n+1,m}^{*} & 0 & 0\\
0 & 0 & \beta_{n+1,m}\beta_{n+1,m}^{*} & 0\\
(\delta_{m})^{2} & 0 & 0 & \beta_{n+1,m}\gamma_{n+1,m}^{*}
\end{pmatrix}_{\overline{(n+1)}\underline{(n+1)}}\otimes C_{n,m} \,.
\end{align*}
%\end{allowdisplaybreaks}

If $N$ is large enough, it becomes a good approximation that there be at most one particle at each discretized lattice site. We then have
\begin{equation}
L_{i}(\lambda_{1},\cdots,\lambda_{n})\left|0\right\rangle _{i}=C_{n,0}\left|0\right\rangle _{i}+C_{n,1}\left|1\right\rangle _{i}
\end{equation}
and we obtain a MPS of the discretized Lieb-Liniger model with OBC given as
\begin{multline}
\left|\lambda_{1},\lambda_{2},\cdots,\lambda_{n}\right\rangle =
\sum_{\{i_{1},i_{2},\cdots,i_{n}\}}
\mathrm{Tr}_{(\bar{V} \otimes \ubar{V})^{\otimes n}}
\Big[\left(C_{n,0}\right)^{i_{1}-1}   
C_{n,1} \left(C_{n,0}\right)^{i_{2}-i_{1}-1}C_{n,1}\cdots
\left(C_{n,0}\right)^{i_{n}-i_{n-1}-1}      \\
C_{n,1}   
\left(C_{n,0}\right)^{N-i_{n}}             
 \mathcal{Q}_{n}\Big]  
\left|x_{1},x_{2},\cdots,x_{n}\right\rangle\,, 
\end{multline}
where $1\leq i_{1}<i_{2}<\cdots<i_{n}\leq N$, $x_{m}$ is related to $i_{m}$ by $x_{m}=i_{m} a$, $\left|x_{1},x_{2},\cdots,x_{n}\right\rangle $ denotes a configuration with particles at (discrete) positions $(x_{1},x_{2},\cdots,x_{n})$, the subscript $n$ indicates the total number of particles, and $\mathcal{Q}_{n}$ is again given by Eq.~(\ref{eq:Qmat3}).

It is clear that we cannot derive a cMPS similar to the ones in the literature~\cite{VerstraeteCirac2010,Chung2015,Chung2016,Ganalh2016}, because we cannot find a basis in which the matrices have a simple form. In~\cite{Maruyama2010}, the authors derived a cMPS for the Lieb-Liniger model with PBC by finding a basis in which the matrices are considerably simpler. Achieving the same for OBC is an interesting and challenging open problem. 
%\end{widetext}
\twocolumngrid                                                                                                                                     %Two columns
%\bibliography{References}

\begin{thebibliography}{66}%
\makeatletter
\providecommand \@ifxundefined [1]{%
 \ifx #1\undefined \expandafter \@firstoftwo
 \else \expandafter \@secondoftwo
\fi
}%
\providecommand \@ifnum [1]{%
 \ifnum #1\expandafter \@firstoftwo
 \else \expandafter \@secondoftwo
\fi
}%
\providecommand \enquote [1]{``#1''}%
\providecommand \bibnamefont  [1]{#1}%
\providecommand \bibfnamefont [1]{#1}%
\providecommand \citenamefont [1]{#1}%
\providecommand\href[0]{\@sanitize\@href}%
\providecommand\@href[1]{\endgroup\@@startlink{#1}\endgroup\@@href}%
\providecommand\@@href[1]{#1\@@endlink}%
\providecommand \@sanitize [0]{\begingroup\catcode`\&12\catcode`\#12\relax}%
\@ifxundefined \pdfoutput {\@firstoftwo}{%
 \@ifnum{\z@=\pdfoutput}{\@firstoftwo}{\@secondoftwo}%
}{%
 \providecommand\@@startlink[1]{\leavevmode}%
 \providecommand\@@endlink[0]{}%
}{%
 \providecommand\@@startlink[1]{%
  \leavevmode
  \pdfstartlink
   attr{/Border[0 0 1 ]/H/I/C[0 1 1]}%
   user{/Subtype/Link/A<</Type/Action/S/URI/URI(#1)>>}%
  \relax
 }%
 \providecommand\@@endlink[0]{\pdfendlink}%
}%
\providecommand \url  [0]{\begingroup\@sanitize \@url }%
\providecommand \@url [1]{\endgroup\@href {#1}{\urlprefix}}%
\providecommand \urlprefix [0]{URL }%
\providecommand \Eprint[0]{\href }%
\@ifxundefined \urlstyle {%
  \providecommand \doi [1]{doi:\discretionary{}{}{}#1}%
}{%
  \providecommand \doi [0]{doi:\discretionary{}{}{}\begingroup
  \urlstyle{rm}\Url }%
}%
\providecommand \doibase [0]{http://dx.doi.org/}%
\providecommand \Doi[1]{\href{\doibase#1}}%
\providecommand \bibAnnote [3]{%
  \BibitemShut{#1}%
  \begin{quotation}\noindent
    \textsc{Key:}\ #2\\\textsc{Annotation:}\ #3%
  \end{quotation}%
}%
\providecommand \bibAnnoteFile [2]{%
  \IfFileExists{#2}{\bibAnnote {#1} {#2} {\input{#2}}}{}%
}%
\providecommand \typeout [0]{\immediate \write \m@ne }%
\providecommand \selectlanguage [0]{\@gobble}%
\providecommand \bibinfo [0]{\@secondoftwo}%
\providecommand \bibfield [0]{\@secondoftwo}%
\providecommand \translation [1]{[#1]}%
\providecommand \BibitemOpen[0]{}%
\providecommand \bibitemStop [0]{}%
\providecommand \bibitemNoStop [0]{.\EOS\space}%
\providecommand \EOS [0]{\spacefactor3000\relax}%
\providecommand \BibitemShut [1]{\csname bibitem#1\endcsname}%
%</preamble>
\bibitem{Bethe1931}%
  \BibitemOpen
  \bibfield{author}{%
  \bibinfo {author} {\bibfnamefont{H.~A.}\ \bibnamefont{Bethe}},\ }%
  \bibfield{journal}{%
  \Doi{10.1007/BF01341708}{\bibinfo {journal} {Z. Phys.}}\ }%
  \textbf{\bibinfo {volume} {71}},\ \bibinfo {pages} {205} (\bibinfo {year}
  {1931})%
  \bibAnnoteFile{NoStop}{Bethe1931}%
\bibitem{Bernal1933}%
  \BibitemOpen
  \bibfield{author}{%
  \bibinfo {author} {\bibfnamefont{J.~D.}\ \bibnamefont{Bernal}}\ and\ \bibinfo
  {author} {\bibfnamefont{R.~H.}\ \bibnamefont{Fowler}},\ }%
  \bibfield{journal}{%
  \Doi{http://dx.doi.org/10.1063/1.1749327}{\bibinfo {journal} {J. Chem.
  Phys.}}\ }%
  \textbf{\bibinfo {volume} {1}},\ \bibinfo {pages} {515} (\bibinfo {year}
  {1933})%
  \bibAnnoteFile{NoStop}{Bernal1933}%
\bibitem{Pauling1935}%
  \BibitemOpen
  \bibfield{author}{%
  \bibinfo {author} {\bibfnamefont{L.}~\bibnamefont{Pauling}},\ }%
  \bibfield{journal}{%
  \Doi{10.1021/ja01315a102}{\bibinfo {journal} {J. Am. Chem. Soc.}}\ }%
  \textbf{\bibinfo {volume} {57}},\ \bibinfo {pages} {2680} (\bibinfo {year}
  {1935})%
  \bibAnnoteFile{NoStop}{Pauling1935}%
\bibitem{Lieb1967a}%
  \BibitemOpen
  \bibfield{author}{%
  \bibinfo {author} {\bibfnamefont{E.~H.}\ \bibnamefont{Lieb}},\ }%
  \bibfield{journal}{%
  \Doi{10.1103/PhysRevLett.18.692}{\bibinfo {journal} {Phys. Rev. Lett.}}\ }%
  \textbf{\bibinfo {volume} {18}},\ \bibinfo {pages} {692} (\bibinfo {year}
  {1967})%
  \bibAnnoteFile{NoStop}{Lieb1967a}%
\bibitem{Lieb1967b}%
  \BibitemOpen
  \bibfield{author}{%
  \bibinfo {author} {\bibfnamefont{E.~H.}\ \bibnamefont{Lieb}},\ }%
  \bibfield{journal}{%
  \Doi{10.1103/PhysRev.162.162}{\bibinfo {journal} {Phys. Rev.}}\ }%
  \textbf{\bibinfo {volume} {162}},\ \bibinfo {pages} {162} (\bibinfo {year}
  {1967})%
  \bibAnnoteFile{NoStop}{Lieb1967b}%
\bibitem{Sutherland1967}%
  \BibitemOpen
  \bibfield{author}{%
  \bibinfo {author} {\bibfnamefont{B.}~\bibnamefont{Sutherland}},\ }%
  \bibfield{journal}{%
  \Doi{10.1103/PhysRevLett.19.103}{\bibinfo {journal} {Phys. Rev. Lett.}}\ }%
  \textbf{\bibinfo {volume} {19}},\ \bibinfo {pages} {103} (\bibinfo {year}
  {1967})%
  \bibAnnoteFile{NoStop}{Sutherland1967}%
\bibitem{Baxter}%
  \BibitemOpen
  \bibfield{author}{%
  \bibinfo {author} {\bibfnamefont{R.~J.}\ \bibnamefont{Baxter}},\ }%
  \emph{\bibinfo {title} {Exactly solved models in statistical mechanics}}\
  (\bibinfo {publisher} {Academic Press},\ \bibinfo {address} {London UK},\
  \bibinfo {year} {1982})%
  \bibAnnoteFile{NoStop}{Baxter}%
\bibitem{White1992}%
  \BibitemOpen
  \bibfield{author}{%
  \bibinfo {author} {\bibfnamefont{S.~R.}\ \bibnamefont{White}},\ }%
  \bibfield{journal}{%
  \Doi{10.1103/PhysRevLett.69.2863}{\bibinfo {journal} {Phys. Rev. Lett.}}\ }%
  \textbf{\bibinfo {volume} {69}},\ \bibinfo {pages} {2863} (\bibinfo {year}
  {1992})%
  \bibAnnoteFile{NoStop}{White1992}%
\bibitem{White1993}%
  \BibitemOpen
  \bibfield{author}{%
  \bibinfo {author} {\bibfnamefont{S.~R.}\ \bibnamefont{White}},\ }%
  \bibfield{journal}{%
  \Doi{10.1103/PhysRevB.48.10345}{\bibinfo {journal} {Phys. Rev. B}}\ }%
  \textbf{\bibinfo {volume} {48}},\ \bibinfo {pages} {10345} (\bibinfo {year}
  {1993})%
  \bibAnnoteFile{NoStop}{White1993}%
\bibitem{Nishino1996}%
  \BibitemOpen
  \bibfield{author}{%
  \bibinfo {author} {\bibfnamefont{T.}~\bibnamefont{Nishino}}\ and\ \bibinfo
  {author} {\bibfnamefont{K.}~\bibnamefont{Okunishi}},\ }%
  \bibfield{journal}{%
  \Doi{10.1143/JPSJ.65.891}{\bibinfo {journal} {J. Phys. Soc. of Jpn.}}\ }%
  \textbf{\bibinfo {volume} {65}},\ \bibinfo {pages} {891} (\bibinfo {year}
  {1996})%
  \bibAnnoteFile{NoStop}{Nishino1996}%
\bibitem{Ostlund1995}%
  \BibitemOpen
  \bibfield{author}{%
  \bibinfo {author} {\bibfnamefont{S.}~\bibnamefont{\"Ostlund}}\ and\ \bibinfo
  {author} {\bibfnamefont{S.}~\bibnamefont{Rommer}},\ }%
  \bibfield{journal}{%
  \Doi{10.1103/PhysRevLett.75.3537}{\bibinfo {journal} {Phys. Rev. Lett.}}\ }%
  \textbf{\bibinfo {volume} {75}},\ \bibinfo {pages} {3537} (\bibinfo {year}
  {1995})%
  \bibAnnoteFile{NoStop}{Ostlund1995}%
\bibitem{VerstraeteMurgCirac2008}%
  \BibitemOpen
  \bibfield{author}{%
  \bibinfo {author} {\bibfnamefont{F.}~\bibnamefont{Verstraete}}, \bibinfo
  {author} {\bibfnamefont{V.}~\bibnamefont{Murg}},\ and\ \bibinfo {author}
  {\bibfnamefont{J.~I.}\ \bibnamefont{Cirac}},\ }%
  \bibfield{journal}{%
  \Doi{10.1080/14789940801912366}{\bibinfo {journal} {Adv. Phys.}}\ }%
  \textbf{\bibinfo {volume} {57}},\ \bibinfo {pages} {143} (\bibinfo {year}
  {2008})%
  \bibAnnoteFile{NoStop}{VerstraeteMurgCirac2008}%
\bibitem{garcia2004}%
  \BibitemOpen
  \bibfield{author}{%
  \bibinfo {author} {\bibfnamefont{J.~J.}\ \bibnamefont{Garc\'{\i}a-Ripoll}},
  \bibinfo {author} {\bibfnamefont{M.~A.}\ \bibnamefont{Martin-Delgado}},\ and\
  \bibinfo {author} {\bibfnamefont{J.~I.}\ \bibnamefont{Cirac}},\ }%
  \bibfield{journal}{%
  \Doi{10.1103/PhysRevLett.93.250405}{\bibinfo {journal} {Phys. Rev. Lett.}}\
  }%
  \textbf{\bibinfo {volume} {93}},\ \bibinfo {pages} {250405} (\bibinfo {year}
  {2004})%
  \bibAnnoteFile{NoStop}{garcia2004}%
\bibitem{verstraete2004}%
  \BibitemOpen
  \bibfield{author}{%
  \bibinfo {author} {\bibfnamefont{F.}~\bibnamefont{Verstraete}}, \bibinfo
  {author} {\bibfnamefont{M.~A.}\ \bibnamefont{Mart\'{\i}n-Delgado}},\ and\
  \bibinfo {author} {\bibfnamefont{J.~I.}\ \bibnamefont{Cirac}},\ }%
  \bibfield{journal}{%
  \Doi{10.1103/PhysRevLett.92.087201}{\bibinfo {journal} {Phys. Rev. Lett.}}\
  }%
  \textbf{\bibinfo {volume} {92}},\ \bibinfo {pages} {087201} (\bibinfo {year}
  {2004})%
  \bibAnnoteFile{NoStop}{verstraete2004}%
\bibitem{popp2005}%
  \BibitemOpen
  \bibfield{author}{%
  \bibinfo {author} {\bibfnamefont{M.}~\bibnamefont{Popp}}, \bibinfo {author}
  {\bibfnamefont{F.}~\bibnamefont{Verstraete}}, \bibinfo {author}
  {\bibfnamefont{M.~A.}\ \bibnamefont{Mart\'{\i}n-Delgado}},\ and\ \bibinfo
  {author} {\bibfnamefont{J.~I.}\ \bibnamefont{Cirac}},\ }%
  \bibfield{journal}{%
  \Doi{10.1103/PhysRevA.71.042306}{\bibinfo {journal} {Phys. Rev. A}}\ }%
  \textbf{\bibinfo {volume} {71}},\ \bibinfo {pages} {042306} (\bibinfo {year}
  {2005})%
  \bibAnnoteFile{NoStop}{popp2005}%
\bibitem{Affleck1987}%
  \BibitemOpen
  \bibfield{author}{%
  \bibinfo {author} {\bibfnamefont{I.}~\bibnamefont{Affleck}}, \bibinfo
  {author} {\bibfnamefont{T.}~\bibnamefont{Kennedy}}, \bibinfo {author}
  {\bibfnamefont{E.~H.}\ \bibnamefont{Lieb}},\ and\ \bibinfo {author}
  {\bibfnamefont{H.}~\bibnamefont{Tasaki}},\ }%
  \bibfield{journal}{%
  \Doi{10.1103/PhysRevLett.59.799}{\bibinfo {journal} {Phys. Rev. Lett.}}\ }%
  \textbf{\bibinfo {volume} {59}},\ \bibinfo {pages} {799} (\bibinfo {year}
  {1987})%
  \bibAnnoteFile{NoStop}{Affleck1987}%
\bibitem{Schollwock2011}%
  \BibitemOpen
  \bibfield{author}{%
  \bibinfo {author} {\bibfnamefont{U.}~\bibnamefont{Schollw\"ock}},\ }%
  \bibfield{journal}{%
  \Doi{http://dx.doi.org/10.1016/j.aop.2010.09.012}{\bibinfo {journal} {Ann.
  Phys.}}\ }%
  \textbf{\bibinfo {volume} {326}},\ \bibinfo {pages} {96 } (\bibinfo {year}
  {2011})%
  \bibAnnoteFile{NoStop}{Schollwock2011}%
\bibitem{SchollwockRevModPhys2005}%
  \BibitemOpen
  \bibfield{author}{%
  \bibinfo {author} {\bibfnamefont{U.}~\bibnamefont{Schollw\"ock}},\ }%
  \bibfield{journal}{%
  \Doi{10.1103/RevModPhys.77.259}{\bibinfo {journal} {Rev. Mod. Phys.}}\ }%
  \textbf{\bibinfo {volume} {77}},\ \bibinfo {pages} {259} (\bibinfo {year}
  {2005})%
  \bibAnnoteFile{NoStop}{SchollwockRevModPhys2005}%
\bibitem{Orus2014}%
  \BibitemOpen
  \bibfield{author}{%
  \bibinfo {author} {\bibfnamefont{R.}~\bibnamefont{Or\'us}},\ }%
  \bibfield{journal}{%
  \Doi{http://dx.doi.org/10.1016/j.aop.2014.06.013}{\bibinfo {journal} {Ann.
  Phys.}}\ }%
  \textbf{\bibinfo {volume} {349}},\ \bibinfo {pages} {117 } (\bibinfo {year}
  {2014})%
  \bibAnnoteFile{NoStop}{Orus2014}%
\bibitem{Alcaraz2004}%
  \BibitemOpen
  \bibfield{author}{%
  \bibinfo {author} {\bibfnamefont{F.~C.}\ \bibnamefont{Alcaraz}}\ and\
  \bibinfo {author} {\bibfnamefont{M.~J.}\ \bibnamefont{Lazo}},\ }%
  \bibfield{journal}{%
  \Doi{10.1088/0305-4470/37/14/001}{\bibinfo {journal} {J. Phys. A}}\ }%
  \textbf{\bibinfo {volume} {37}},\ \bibinfo {pages} {4149} (\bibinfo {year}
  {2004})%
  \bibAnnoteFile{NoStop}{Alcaraz2004}%
\bibitem{Alcaraz2006}%
  \BibitemOpen
  \bibfield{author}{%
  \bibinfo {author} {\bibfnamefont{F.~C.}\ \bibnamefont{Alcaraz}}\ and\
  \bibinfo {author} {\bibfnamefont{M.~J.}\ \bibnamefont{Lazo}},\ }%
  \bibfield{journal}{%
  \Doi{10.1088/0305-4470/39/36/014}{\bibinfo {journal} {J. Phys. A}}\ }%
  \textbf{\bibinfo {volume} {39}},\ \bibinfo {pages} {11335} (\bibinfo {year}
  {2006})%
  \bibAnnoteFile{NoStop}{Alcaraz2006}%
\bibitem{Derrida1993}%
  \BibitemOpen
  \bibfield{author}{%
  \bibinfo {author} {\bibfnamefont{B.}~\bibnamefont{Derrida}}, \bibinfo
  {author} {\bibfnamefont{M.~R.}\ \bibnamefont{Evans}}, \bibinfo {author}
  {\bibfnamefont{V.}~\bibnamefont{Hakim}},\ and\ \bibinfo {author}
  {\bibfnamefont{V.}~\bibnamefont{Pasquier}},\ }%
  \bibfield{journal}{%
  \Doi{10.1088/0305-4470/26/7/011}{\bibinfo {journal} {J. Phys. A}}\ }%
  \textbf{\bibinfo {volume} {26}},\ \bibinfo {pages} {1493} (\bibinfo {year}
  {1993})%
  \bibAnnoteFile{NoStop}{Derrida1993}%
\bibitem{Aneva2016}%
  \BibitemOpen
  \bibfield{author}{%
  \bibinfo {author} {\bibfnamefont{B.~L.}\ \bibnamefont{Aneva}}\ and\ \bibinfo
  {author} {\bibfnamefont{J.~G.}\ \bibnamefont{Brankov}},\ }%
  \bibfield{journal}{%
  \Doi{10.1103/PhysRevE.94.022138}{\bibinfo {journal} {Phys. Rev. E}}\ }%
  \textbf{\bibinfo {volume} {94}},\ \bibinfo {pages} {022138} (\bibinfo {year}
  {2016})%
  \bibAnnoteFile{NoStop}{Aneva2016}%
\bibitem{Crampe2016}%
  \BibitemOpen
  \bibfield{author}{%
  \bibinfo {author} {\bibfnamefont{N.}~\bibnamefont{Crampe}}, \bibinfo {author}
  {\bibfnamefont{E.}~\bibnamefont{Ragoucy}}, \bibinfo {author}
  {\bibfnamefont{V.}~\bibnamefont{Rittenberg}},\ and\ \bibinfo {author}
  {\bibfnamefont{M.}~\bibnamefont{Vanicat}},\ }%
  \bibfield{journal}{%
  \Doi{10.1103/PhysRevE.94.032102}{\bibinfo {journal} {Phys. Rev. E}}\ }%
  \textbf{\bibinfo {volume} {94}},\ \bibinfo {pages} {032102} (\bibinfo {year}
  {2016})%
  \bibAnnoteFile{NoStop}{Crampe2016}%
\bibitem{GolinelliMallick2006}%
  \BibitemOpen
  \bibfield{author}{%
  \bibinfo {author} {\bibfnamefont{O.}~\bibnamefont{Golinelli}}\ and\ \bibinfo
  {author} {\bibfnamefont{K.}~\bibnamefont{Mallick}},\ }%
  \bibfield{journal}{%
  \Doi{10.1088/0305-4470/39/34/004}{\bibinfo {journal} {J. Phys. A}}\ }%
  \textbf{\bibinfo {volume} {39}},\ \bibinfo {pages} {10647} (\bibinfo {year}
  {2006})%
  \bibAnnoteFile{NoStop}{GolinelliMallick2006}%
\bibitem{Nepomechie1999}%
  \BibitemOpen
  \bibfield{author}{%
  \bibinfo {author} {\bibfnamefont{R.~I.}\ \bibnamefont{Nepomechie}},\ }%
  \bibfield{journal}{%
  \Doi{10.1142/S0217979299002800}{\bibinfo {journal} {Int. J. Mod. Phys. B}}\
  }%
  \textbf{\bibinfo {volume} {13}},\ \bibinfo {pages} {2973} (\bibinfo {year}
  {1999})%
  \bibAnnoteFile{NoStop}{Nepomechie1999}%
\bibitem{Korepin1993}%
  \BibitemOpen
  \bibfield{author}{%
  \bibinfo {author} {\bibfnamefont{V.~E.}\ \bibnamefont{Korepin}}, \bibinfo
  {author} {\bibfnamefont{N.~M.}\ \bibnamefont{Bogoliubov}},\ and\ \bibinfo
  {author} {\bibfnamefont{A.~G.}\ \bibnamefont{Izergin}},\ }%
  {\emph{\bibinfo {title} {Quantum Inverse
  Scattering Method and Correlation Functions}}}\ (\bibinfo {publisher}
  {Cambridge University Press},\ \bibinfo {year} {1993})%
  \bibAnnoteFile{NoStop}{Korepin1993}%
\bibitem{Katsura2010}%
  \BibitemOpen
  \bibfield{author}{%
  \bibinfo {author} {\bibfnamefont{H.}~\bibnamefont{Katsura}}\ and\ \bibinfo
  {author} {\bibfnamefont{I.}~\bibnamefont{Maruyama}},\ }%
  \bibfield{journal}{%
  \Doi{10.1088/1751-8113/43/17/175003}{\bibinfo {journal} {J. Phys. A}}\ }%
  \textbf{\bibinfo {volume} {43}},\ \bibinfo {pages} {175003} (\bibinfo {year}
  {2010})%
  \bibAnnoteFile{NoStop}{Katsura2010}%
\bibitem{LiebLiniger1963}%
  \BibitemOpen
  \bibfield{author}{%
  \bibinfo {author} {\bibfnamefont{E.~H.}\ \bibnamefont{Lieb}}\ and\ \bibinfo
  {author} {\bibfnamefont{W.}~\bibnamefont{Liniger}},\ }%
  \bibfield{journal}{%
  \Doi{10.1103/PhysRev.130.1605}{\bibinfo {journal} {Phys. Rev.}}\ }%
  \textbf{\bibinfo {volume} {130}},\ \bibinfo {pages} {1605} (\bibinfo {year}
  {1963})%
  \bibAnnoteFile{NoStop}{LiebLiniger1963}%
\bibitem{Maruyama2010}%
  \BibitemOpen
  \bibfield{author}{%
  \bibinfo {author} {\bibfnamefont{I.}~\bibnamefont{Maruyama}}\ and\ \bibinfo
  {author} {\bibfnamefont{H.}~\bibnamefont{Katsura}},\ }%
  \bibfield{journal}{%
  \Doi{10.1143/JPSJ.79.073002}{\bibinfo {journal} {J. Phys. Soc. Jpn.}}\ }%
  \textbf{\bibinfo {volume} {79}},\ \bibinfo {pages} {073002} (\bibinfo {year}
  {2010})%
  \bibAnnoteFile{NoStop}{Maruyama2010}%
\bibitem{Korepin2000}%
  \BibitemOpen
  \bibfield{author}{%
  \bibinfo {author} {\bibfnamefont{V.}~\bibnamefont{Korepin}}\ and\ \bibinfo
  {author} {\bibfnamefont{P.}~\bibnamefont{Zinn-Justin}},\ }%
  \bibfield{journal}{%
  \Doi{10.1088/0305-4470/33/40/304}{\bibinfo {journal} {J. Phys. A}}\ }%
  \textbf{\bibinfo {volume} {33}},\ \bibinfo {pages} {7053} (\bibinfo {year}
  {2000})%
  \bibAnnoteFile{NoStop}{Korepin2000}%
\bibitem{ZinnJustin2000}%
  \BibitemOpen
  \bibfield{author}{%
  \bibinfo {author} {\bibfnamefont{P.}~\bibnamefont{Zinn-Justin}},\ }%
  \bibfield{journal}{%
  \Doi{10.1103/PhysRevE.62.3411}{\bibinfo {journal} {Phys. Rev. E}}\ }%
  \textbf{\bibinfo {volume} {62}},\ \bibinfo {pages} {3411} (\bibinfo {year}
  {2000})%
  \bibAnnoteFile{NoStop}{ZinnJustin2000}%
\bibitem{Cugliandolo2015}%
  \BibitemOpen
  \bibfield{author}{%
  \bibinfo {author} {\bibfnamefont{L.~F.}\ \bibnamefont{Cugliandolo}}, \bibinfo
  {author} {\bibfnamefont{G.}~\bibnamefont{Gonnella}},\ and\ \bibinfo {author}
  {\bibfnamefont{A.}~\bibnamefont{Pelizzola}},\ }%
  \bibfield{journal}{%
  \Doi{10.1088/1742-5468/2015/06/P06008}{\bibinfo {journal} {J. Stat. Mech.:
  Th. and Exp.}}\ }%
  \textbf{\bibinfo {volume} {2015}},\ \bibinfo {pages} {P06008} (\bibinfo
  {year} {2015})%
  \bibAnnoteFile{NoStop}{Cugliandolo2015}%
\bibitem{Brascamp1973}%
  \BibitemOpen
  \bibfield{author}{%
  \bibinfo {author} {\bibfnamefont{H.~J.}\ \bibnamefont{Brascamp}}, \bibinfo
  {author} {\bibfnamefont{H.}~\bibnamefont{Kunz}},\ and\ \bibinfo {author}
  {\bibfnamefont{F.~Y.}\ \bibnamefont{Wu}},\ }%
  \bibfield{journal}{%
  \Doi{http://dx.doi.org/10.1063/1.1666271}{\bibinfo {journal} {J. Math.
  Phys.}}\ }%
  \textbf{\bibinfo {volume} {14}},\ \bibinfo {pages} {1927} (\bibinfo {year}
  {1973})%
  \bibAnnoteFile{NoStop}{Brascamp1973}%
\bibitem{VerstraetePorrasCirac2004}%
  \BibitemOpen
  \bibfield{author}{%
  \bibinfo {author} {\bibfnamefont{F.}~\bibnamefont{Verstraete}}, \bibinfo
  {author} {\bibfnamefont{D.}~\bibnamefont{Porras}},\ and\ \bibinfo {author}
  {\bibfnamefont{J.~I.}\ \bibnamefont{Cirac}},\ }%
  \bibfield{journal}{%
  \Doi{10.1103/PhysRevLett.93.227205}{\bibinfo {journal} {Phys. Rev. Lett.}}\
  }%
  \textbf{\bibinfo {volume} {93}},\ \bibinfo {pages} {227205} (\bibinfo {year}
  {2004})%
  \bibAnnoteFile{NoStop}{VerstraetePorrasCirac2004}%
\bibitem{VerstraeteCirac2010}%
  \BibitemOpen
  \bibfield{author}{%
  \bibinfo {author} {\bibfnamefont{F.}~\bibnamefont{Verstraete}}\ and\ \bibinfo
  {author} {\bibfnamefont{J.~I.}\ \bibnamefont{Cirac}},\ }%
  \bibfield{journal}{%
  \Doi{10.1103/PhysRevLett.104.190405}{\bibinfo {journal} {Phys. Rev. Lett.}}\
  }%
  \textbf{\bibinfo {volume} {104}},\ \bibinfo {pages} {190405} (\bibinfo {year}
  {2010})%
  \bibAnnoteFile{NoStop}{VerstraeteCirac2010}%
\bibitem{Sklyanin1988}%
  \BibitemOpen
  \bibfield{author}{%
  \bibinfo {author} {\bibfnamefont{E.~K.}\ \bibnamefont{Sklyanin}},\ }%
  \bibfield{journal}{%
  \Doi{10.1088/0305-4470/21/10/015}{\bibinfo {journal} {J. Phys. A}}\ }%
  \textbf{\bibinfo {volume} {21}},\ \bibinfo {pages} {2375} (\bibinfo {year}
  {1988})%
  \bibAnnoteFile{NoStop}{Sklyanin1988}%
\bibitem{Abad1996}%
  \BibitemOpen
  \bibfield{author}{%
  \bibinfo {author} {\bibfnamefont{J.}~\bibnamefont{Abad}}\ and\ \bibinfo
  {author} {\bibfnamefont{M.}~\bibnamefont{Rios}},\ }%
  \bibfield{journal}{%
  \Doi{10.1103/PhysRevB.53.14000}{\bibinfo {journal} {Phys. Rev. B}}\ }%
  \textbf{\bibinfo {volume} {53}},\ \bibinfo {pages} {14000} (\bibinfo {year}
  {1996})%
  \bibAnnoteFile{NoStop}{Abad1996}%
\bibitem{Bolech2002}%
  \BibitemOpen
  \bibfield{author}{%
  \bibinfo {author} {\bibfnamefont{C.~J.}\ \bibnamefont{Bolech}}\ and\ \bibinfo
  {author} {\bibfnamefont{N.}~\bibnamefont{Andrei}},\ }%
  \bibfield{journal}{%
  \Doi{10.1103/PhysRevLett.88.237206}{\bibinfo {journal} {Phys. Rev. Lett.}}\
  }%
  \textbf{\bibinfo {volume} {88}},\ \bibinfo {pages} {237206} (\bibinfo {year}
  {2002})%
  \bibAnnoteFile{NoStop}{Bolech2002}%
\bibitem{Bolech2005}%
  \BibitemOpen
  \bibfield{author}{%
  \bibinfo {author} {\bibfnamefont{C.~J.}\ \bibnamefont{Bolech}}\ and\ \bibinfo
  {author} {\bibfnamefont{N.}~\bibnamefont{Andrei}},\ }%
  \bibfield{journal}{%
  \Doi{10.1103/PhysRevB.71.205104}{\bibinfo {journal} {Phys. Rev. B}}\ }%
  \textbf{\bibinfo {volume} {71}},\ \bibinfo {pages} {205104} (\bibinfo {year}
  {2005})%
  \bibAnnoteFile{NoStop}{Bolech2005}%
\bibitem{Draxler2017}%
  \BibitemOpen
  \bibfield{author}{%
  \bibinfo {author} {\bibfnamefont{D.}~\bibnamefont{Draxler}}, \bibinfo
  {author} {\bibfnamefont{J.}~\bibnamefont{Haegeman}}, \bibinfo {author}
  {\bibfnamefont{F.}~\bibnamefont{Verstraete}},\ and\ \bibinfo {author}
  {\bibfnamefont{M.}~\bibnamefont{Rizzi}},\ }%
  \bibfield{journal}{%
  \Doi{10.1103/PhysRevB.95.045145}{\bibinfo {journal} {Phys. Rev. B}}\ }%
  \textbf{\bibinfo {volume} {95}},\ \bibinfo {pages} {045145} (\bibinfo {year}
  {2017})%
  \bibAnnoteFile{NoStop}{Draxler2017}%
\bibitem{MurgKorepin2012}%
  \BibitemOpen
  \bibfield{author}{%
  \bibinfo {author} {\bibfnamefont{V.}~\bibnamefont{Murg}}, \bibinfo {author}
  {\bibfnamefont{V.~E.}\ \bibnamefont{Korepin}},\ and\ \bibinfo {author}
  {\bibfnamefont{F.}~\bibnamefont{Verstraete}},\ }%
  \bibfield{journal}{%
  \Doi{10.1103/PhysRevB.86.045125}{\bibinfo {journal} {Phys. Rev. B}}\ }%
  \textbf{\bibinfo {volume} {86}},\ \bibinfo {pages} {045125} (\bibinfo {year}
  {2012})%
  \bibAnnoteFile{NoStop}{MurgKorepin2012}%
\bibitem{McCulloch2008}%
  \BibitemOpen
  \bibfield{author}{%
  \bibinfo {author} {\bibfnamefont{I.~P.}\ \bibnamefont{{McCulloch}}},\ }%
  \bibfield{journal}{%
  \bibinfo {journal} {arXiv:0804.2509~}}%
   (\bibinfo {year} {2008})%
  \bibAnnoteFile{NoStop}{McCulloch2008}%
\bibitem{Kjall2013}%
  \BibitemOpen
  \bibfield{author}{%
  \bibinfo {author} {\bibfnamefont{J.~A.}\ \bibnamefont{Kj\"all}}, \bibinfo
  {author} {\bibfnamefont{M.~P.}\ \bibnamefont{Zaletel}}, \bibinfo {author}
  {\bibfnamefont{R.~S.~K.}\ \bibnamefont{Mong}}, \bibinfo {author}
  {\bibfnamefont{J.~H.}\ \bibnamefont{Bardarson}},\ and\ \bibinfo {author}
  {\bibfnamefont{F.}~\bibnamefont{Pollmann}},\ }%
  \bibfield{journal}{%
  \Doi{10.1103/PhysRevB.87.235106}{\bibinfo {journal} {Phys. Rev. B}}\ }%
  \textbf{\bibinfo {volume} {87}},\ \bibinfo {pages} {235106} (\bibinfo {year}
  {2013})%
  \bibAnnoteFile{NoStop}{Kjall2013}%
\newpage
\bibitem{Vidal2007}%
  \BibitemOpen
  \bibfield{author}{%
  \bibinfo {author} {\bibfnamefont{G.}~\bibnamefont{Vidal}},\ }%
  \bibfield{journal}{%
  \Doi{10.1103/PhysRevLett.98.070201}{\bibinfo {journal} {Phys. Rev. Lett.}}\
  }%
  \textbf{\bibinfo {volume} {98}},\ \bibinfo {pages} {070201} (\bibinfo {year}
  {2007})%
  \bibAnnoteFile{NoStop}{Vidal2007}%
\bibitem{OrusVidal2008}%
  \BibitemOpen
  \bibfield{author}{%
  \bibinfo {author} {\bibfnamefont{R.}~\bibnamefont{Or\'us}}\ and\ \bibinfo
  {author} {\bibfnamefont{G.}~\bibnamefont{Vidal}},\ }%
  \bibfield{journal}{%
  \Doi{10.1103/PhysRevB.78.155117}{\bibinfo {journal} {Phys. Rev. B}}\ }%
  \textbf{\bibinfo {volume} {78}},\ \bibinfo {pages} {155117} (\bibinfo {year}
  {2008})%
  \bibAnnoteFile{NoStop}{OrusVidal2008}%
\bibitem{Chung2015}%
  \BibitemOpen
  \bibfield{author}{%
  \bibinfo {author} {\bibfnamefont{S.~S.}\ \bibnamefont{Chung}}, \bibinfo
  {author} {\bibfnamefont{K.}~\bibnamefont{Sun}},\ and\ \bibinfo {author}
  {\bibfnamefont{C.~J.}\ \bibnamefont{Bolech}},\ }%
  \bibfield{journal}{%
  \Doi{10.1103/PhysRevB.91.121108}{\bibinfo {journal} {Phys. Rev. B}}\ }%
  \textbf{\bibinfo {volume} {91}},\ \bibinfo {pages} {121108(R)} (\bibinfo
  {year} {2015})%
  \bibAnnoteFile{NoStop}{Chung2015}%
\bibitem{Kancharla2001}%
  \BibitemOpen
  \bibfield{author}{%
  \bibinfo {author} {\bibfnamefont{S.~S.}\ \bibnamefont{Kancharla}}\ and\
  \bibinfo {author} {\bibfnamefont{C.~J.}\ \bibnamefont{Bolech}},\ }%
  \bibfield{journal}{%
  \Doi{10.1103/PhysRevB.64.085119}{\bibinfo {journal} {Phys. Rev. B}}\ }%
  \textbf{\bibinfo {volume} {64}},\ \bibinfo {pages} {085119} (\bibinfo {year}
  {2001})%
  \bibAnnoteFile{NoStop}{Kancharla2001}%
\bibitem{Phien2012}%
  \BibitemOpen
  \bibfield{author}{%
  \bibinfo {author} {\bibfnamefont{H.~N.}\ \bibnamefont{Phien}}, \bibinfo
  {author} {\bibfnamefont{G.}~\bibnamefont{Vidal}},\ and\ \bibinfo {author}
  {\bibfnamefont{I.~P.}\ \bibnamefont{McCulloch}},\ }%
  \bibfield{journal}{%
  \Doi{10.1103/PhysRevB.86.245107}{\bibinfo {journal} {Phys. Rev. B}}\ }%
  \textbf{\bibinfo {volume} {86}},\ \bibinfo {pages} {245107} (\bibinfo {year}
  {2012})%
  \bibAnnoteFile{NoStop}{Phien2012}%
\bibitem{Tsai2000}%
  \BibitemOpen
  \bibfield{author}{%
  \bibinfo {author} {\bibfnamefont{S.-W.}\ \bibnamefont{Tsai}}\ and\ \bibinfo
  {author} {\bibfnamefont{J.~B.}\ \bibnamefont{Marston}},\ }%
  \bibfield{journal}{%
  \Doi{10.1103/PhysRevB.62.5546}{\bibinfo {journal} {Phys. Rev. B}}\ }%
  \textbf{\bibinfo {volume} {62}},\ \bibinfo {pages} {5546} (\bibinfo {year}
  {2000})%
  \bibAnnoteFile{NoStop}{Tsai2000}%
\bibitem{Lauchli2006}%
  \BibitemOpen
  \bibfield{author}{%
  \bibinfo {author} {\bibfnamefont{A.}~\bibnamefont{L\"auchli}}, \bibinfo
  {author} {\bibfnamefont{G.}~\bibnamefont{Schmid}},\ and\ \bibinfo {author}
  {\bibfnamefont{S.}~\bibnamefont{Trebst}},\ }%
  \bibfield{journal}{%
  \Doi{10.1103/PhysRevB.74.144426}{\bibinfo {journal} {Phys. Rev. B}}\ }%
  \textbf{\bibinfo {volume} {74}},\ \bibinfo {pages} {144426} (\bibinfo {year}
  {2006})%
  \bibAnnoteFile{NoStop}{Lauchli2006}%
\bibitem{Andres2008}%
  \BibitemOpen
  \bibfield{author}{%
  \bibinfo {author} {\bibfnamefont{M.}~\bibnamefont{Andres}}, \bibinfo {author}
  {\bibfnamefont{I.}~\bibnamefont{Schneider}},\ and\ \bibinfo {author}
  {\bibfnamefont{S.}~\bibnamefont{Eggert}},\ }%
  \bibfield{journal}{%
  \Doi{10.1103/PhysRevB.77.014429}{\bibinfo {journal} {Phys. Rev. B}}\ }%
  \textbf{\bibinfo {volume} {77}},\ \bibinfo {pages} {014429} (\bibinfo {year}
  {2008})%
  \bibAnnoteFile{NoStop}{Andres2008}%
\bibitem{Pixley2014}%
  \BibitemOpen
  \bibfield{author}{%
  \bibinfo {author} {\bibfnamefont{J.~H.}\ \bibnamefont{Pixley}}, \bibinfo
  {author} {\bibfnamefont{A.}~\bibnamefont{Shashi}},\ and\ \bibinfo {author}
  {\bibfnamefont{A.~H.}\ \bibnamefont{Nevidomskyy}},\ }%
  \bibfield{journal}{%
  \Doi{10.1103/PhysRevB.90.214426}{\bibinfo {journal} {Phys. Rev. B}}\ }%
  \textbf{\bibinfo {volume} {90}},\ \bibinfo {pages} {214426} (\bibinfo {year}
  {2014})%
  \bibAnnoteFile{NoStop}{Pixley2014}%
\bibitem{Takhtajan1982}%
  \BibitemOpen
  \bibfield{author}{%
  \bibinfo {author} {\bibfnamefont{L.}~\bibnamefont{Takhtajan}},\ }%
  \bibfield{journal}{%
  \Doi{http://dx.doi.org/10.1016/0375-9601(82)90764-2}{\bibinfo {journal}
  {Phys. Lett. A}}\ }%
  \textbf{\bibinfo {volume} {87}},\ \bibinfo {pages} {479 } (\bibinfo {year}
  {1982})%
  \bibAnnoteFile{NoStop}{Takhtajan1982}%
\bibitem{Babujian1982}%
  \BibitemOpen
  \bibfield{author}{%
  \bibinfo {author} {\bibfnamefont{H.}~\bibnamefont{Babujian}},\ }%
  \bibfield{journal}{%
  \Doi{http://dx.doi.org/10.1016/0375-9601(82)90403-0}{\bibinfo {journal}
  {Phys. Lett. A}}\ }%
  \textbf{\bibinfo {volume} {90}},\ \bibinfo {pages} {479 } (\bibinfo {year}
  {1982})%
  \bibAnnoteFile{NoStop}{Babujian1982}%
\bibitem{Lai1974}%
  \BibitemOpen
  \bibfield{author}{%
  \bibinfo {author} {\bibfnamefont{C.~K.}\ \bibnamefont{Lai}},\ }%
  \bibfield{journal}{%
  \Doi{10.1063/1.1666522}{\bibinfo {journal} {J. Math. Phys.}}\ }%
  \textbf{\bibinfo {volume} {15}},\ \bibinfo {pages} {1675} (\bibinfo {year}
  {1974})%
  \bibAnnoteFile{NoStop}{Lai1974}%
\bibitem{Sutherland1975}%
  \BibitemOpen
  \bibfield{author}{%
  \bibinfo {author} {\bibfnamefont{B.}~\bibnamefont{Sutherland}},\ }%
  \bibfield{journal}{%
  \Doi{10.1103/PhysRevB.12.3795}{\bibinfo {journal} {Phys. Rev. B}}\ }%
  \textbf{\bibinfo {volume} {12}},\ \bibinfo {pages} {3795} (\bibinfo {year}
  {1975})%
  \bibAnnoteFile{NoStop}{Sutherland1975}%
\bibitem{Poilblanc1993}%
  \BibitemOpen
  \bibfield{author}{%
  \bibinfo {author} {\bibfnamefont{D.}~\bibnamefont{Poilblanc}}, \bibinfo
  {author} {\bibfnamefont{T.}~\bibnamefont{Ziman}}, \bibinfo {author}
  {\bibfnamefont{J.}~\bibnamefont{Bellissard}}, \bibinfo {author}
  {\bibfnamefont{F.}~\bibnamefont{Mila}},\ and\ \bibinfo {author}
  {\bibfnamefont{G.}~\bibnamefont{Montambaux}},\ }%
  \bibfield{journal}{%
  \Doi{10.1209/0295-5075/22/7/010}{\bibinfo {journal} {Europhys. Lett.}}\ }%
  \textbf{\bibinfo {volume} {22}},\ \bibinfo {pages} {537} (\bibinfo {year}
  {1993})%
  \bibAnnoteFile{NoStop}{Poilblanc1993}%
\bibitem{Narozhny1998}%
  \BibitemOpen
  \bibfield{author}{%
  \bibinfo {author} {\bibfnamefont{B.~N.}\ \bibnamefont{Narozhny}}, \bibinfo
  {author} {\bibfnamefont{A.~J.}\ \bibnamefont{Millis}},\ and\ \bibinfo
  {author} {\bibfnamefont{N.}~\bibnamefont{Andrei}},\ }%
  \bibfield{journal}{%
  \Doi{10.1103/PhysRevB.58.R2921}{\bibinfo {journal} {Phys. Rev. B}}\ }%
  \textbf{\bibinfo {volume} {58}},\ \bibinfo {pages} {R2921} (\bibinfo {year}
  {1998})%
  \bibAnnoteFile{NoStop}{Narozhny1998}%
\bibitem{Wang2015}%
  \BibitemOpen
  \bibfield{author}{%
  \bibinfo {author} {\bibfnamefont{Y.}~\bibnamefont{Wang}}, \bibinfo {author}
  {\bibfnamefont{W.-L.}\ \bibnamefont{Yang}}, \bibinfo {author}
  {\bibfnamefont{J.}~\bibnamefont{Cao}},\ and\ \bibinfo {author}
  {\bibfnamefont{K.}~\bibnamefont{Shi}},\ }%
  {\emph{\bibinfo {title} {Off-Diagonal Bethe
  Ansatz for Exactly Solvable Models}}}\ (\bibinfo {publisher} {Springer-Verlag
  Berlin Heidelberg},\ \bibinfo {year} {2015})%
  \bibAnnoteFile{NoStop}{Wang2015}%
\bibitem{Tarasov1983}%
  \BibitemOpen
  \bibfield{author}{%
  \bibinfo {author} {\bibfnamefont{V.~O.}\ \bibnamefont{Tarasov}}, \bibinfo
  {author} {\bibfnamefont{L.~A.}\ \bibnamefont{Takhtadzhyan}},\ and\ \bibinfo
  {author} {\bibfnamefont{L.~D.}\ \bibnamefont{Faddeev}},\ }%
  \bibfield{journal}{%
  \Doi{10.1007/BF01018648}{\bibinfo {journal} {Theor. Math. Phys.}}\ }%
  \textbf{\bibinfo {volume} {57}},\ \bibinfo {pages} {1059} (\bibinfo {year}
  {1983})%
  \bibAnnoteFile{NoStop}{Tarasov1983}%
\bibitem{Mei2016}%
  \BibitemOpen
  \bibfield{author}{%
  \bibinfo {author} {\bibfnamefont{Z.}~\bibnamefont{Mei}}, \bibinfo {author}
  {\bibfnamefont{L.}~\bibnamefont{Vidmar}}, \bibinfo {author}
  {\bibfnamefont{F.}~\bibnamefont{Heidrich-Meisner}},\ and\ \bibinfo {author}
  {\bibfnamefont{C.~J.}\ \bibnamefont{Bolech}},\ }%
  \bibfield{journal}{%
  \Doi{10.1103/PhysRevA.93.021607}{\bibinfo {journal} {Phys. Rev. A}}\ }%
  \textbf{\bibinfo {volume} {93}},\ \bibinfo {pages} {021607(R)} (\bibinfo
  {year} {2016})%
  \bibAnnoteFile{NoStop}{Mei2016}%
\bibitem{Bolech2012}%
  \BibitemOpen
  \bibfield{author}{%
  \bibinfo {author} {\bibfnamefont{C.~J.}\ \bibnamefont{Bolech}}, \bibinfo
  {author} {\bibfnamefont{F.}~\bibnamefont{Heidrich-Meisner}}, \bibinfo
  {author} {\bibfnamefont{S.}~\bibnamefont{Langer}}, \bibinfo {author}
  {\bibfnamefont{I.~P.}\ \bibnamefont{McCulloch}}, \bibinfo {author}
  {\bibfnamefont{G.}~\bibnamefont{Orso}},\ and\ \bibinfo {author}
  {\bibfnamefont{M.}~\bibnamefont{Rigol}},\ }%
  \bibfield{journal}{%
  \Doi{10.1103/PhysRevLett.109.110602}{\bibinfo {journal} {Phys. Rev. Lett.}}\
  }%
  \textbf{\bibinfo {volume} {109}},\ \bibinfo {pages} {110602} (\bibinfo {year}
  {2012})%
  \bibAnnoteFile{NoStop}{Bolech2012}%
\bibitem{Gaudin1971}%
  \BibitemOpen
  \bibfield{author}{%
  \bibinfo {author} {\bibfnamefont{M.}~\bibnamefont{Gaudin}},\ }%
  \bibfield{journal}{%
  \Doi{10.1103/PhysRevA.4.386}{\bibinfo {journal} {Phys. Rev. A}}\ }%
  \textbf{\bibinfo {volume} {4}},\ \bibinfo {pages} {386} (\bibinfo {year}
  {1971})%
  \bibAnnoteFile{NoStop}{Gaudin1971}%
\bibitem{Chung2016}%
  \BibitemOpen
  \bibfield{author}{%
  \bibinfo {author} {\bibfnamefont{S.~S.}\ \bibnamefont{Chung}}\ and\ \bibinfo
  {author} {\bibfnamefont{C.~J.}\ \bibnamefont{Bolech}},\ }%
  \bibfield{journal}{%
  \bibinfo {journal} {arXiv:1612.03149~}}%
   (\bibinfo {year} {2016})%
  \bibAnnoteFile{NoStop}{Chung2016}%
\bibitem{Ganalh2016}%
  \BibitemOpen
  \bibfield{author}{%
  \bibinfo {author} {\bibfnamefont{M.}~\bibnamefont{Ganahl}}, \bibinfo {author}
  {\bibfnamefont{J.}~\bibnamefont{Rinc\'on}},\ and\ \bibinfo {author}
  {\bibfnamefont{G.}~\bibnamefont{Vidal}},\ }%
  \bibfield{journal}{%
  \bibinfo {journal} {arXiv:1611.03779~}}%
   (\bibinfo {year} {2016})%
  \bibAnnoteFile{NoStop}{Ganalh2016}%
\end{thebibliography}

%Merlin.mbs v4.21 2009-07-09.
%

\end{document}